\documentclass[a4paper,11pt]{article}

\usepackage{jcappub}
\usepackage{graphicx}
\usepackage{natbib}
\usepackage{color}
\usepackage{lmodern}
\usepackage[normalem]{ulem}

\title{Dynamics of galaxies and clusters in \textit{refracted gravity}}

\author[a]{Titos Matsakos}
\author[b,c]{and Antonaldo Diaferio}

\affiliation[a]{
  Dept. of Astronomy \& Astrophysics, The University of Chicago,
  Chicago, IL 60637, USA}
\affiliation[b]{
  Dipartimento di Fisica, Universit\`a di Torino, via P. Giuria 1, 10125,
  Torino, Italy}
\affiliation[c]{
  Istituto Nazionale di Fisica Nucleare (INFN), Sezione di Torino, Torino,
  Italy}

\emailAdd{titos.matsakos@uchicago.edu}
\emailAdd{diaferio@ph.unito.it}

\abstract{
We investigate the proof of concept and the implications of \textit{refracted
gravity}, a novel modified gravity aimed to solve the discrepancy between the
luminous and the dynamical mass of cosmic structures without resorting to dark
matter.
Inspired by the behavior of electric fields in matter, refracted gravity
introduces a gravitational permittivity that depends on the local mass density
and modifies the standard Poisson equation.
The resulting gravitational field can become more intense than the Newtonian
field and can mimic the presence of dark matter.
We show that the refracted gravitational field correctly describes (1) the
rotation curves and the Tully-Fisher relation of disk galaxies; and (2) the
observed temperature profile of the X-ray gas of galaxy clusters.
According to these promising results, we conclude that refracted gravity
deserves further investigation.
}

\keywords{
  modified gravity, rotation curves of galaxies, galaxy dynamics,
  galaxy clusters
}

\arxivnumber{??}

\begin{document}
\maketitle
\flushbottom

\section{Introduction}
  \label{sec:introduction}

It is a well-established fact that visible matter cannot account for the
dynamics observed from galactic scales up to cosmological scales when described
by General Relativity and its Newtonian weak-field limit.
In particular, significant deviations appear in galactic rotation curves, where
the velocities of stars indicate an acceleration or gravitational force
proportional to the inverse of the radial distance, $r^{-1}$, rather than
$r^{-2}$.\footnote{
Hereafter, we will use the terms acceleration and force interchangeably, where
force clearly refers to force per unit mass.}
Moreover, when interpreted in standard gravity, the velocity dispersion of
galaxies in clusters, the temperature of the intracluster gas, and gravitational
lensing effects, all suggest that the gravitating mass of clusters is five to
ten times larger than the mass that emits the electromagnetic radiation.
The most widely accepted interpretation of this mass discrepancy is the
existence of dark matter (DM) which originates the additional gravitational pull
(e.g. \cite{DelPopolo13} for a review).

In this paper we propose a new concept that is inspired by the theory of
electrodynamics in matter.
In particular, a dielectric medium of a non-uniform permittivity can affect both
the direction and magnitude of electric fields.
Here, we investigate whether a similar framework could also describe the
peculiar behavior of gravity in large-scale systems.
We find that the hypothesis of a ``refracted gravitational field'' can in fact
capture many of the observed properties of disk galaxies and galaxy clusters, as
well as other objects.

The paper is structured as follows.
In section~\ref{sec:indications} we introduce the concept of \textit{refracted
gravity} (RG) and we describe its mathematical formulation.
At the end of that section we present a framework that simplifies the
calculations, we derive some of its direct implications, and in
section~\ref{sec:spirals} we apply it to disk galaxies.
In section~\ref{sec:rgpoisson} we solve the full RG equation, apply it to some
specific observed disk galaxies and clusters of galaxies, and confront its
predictions with several mass-discrepancy observations.
In section~\ref{sec:discussion} we compare RG with DM and with MOdified
Newtonian Dynamics (MOND \cite{Milgrom83a, Milgrom83b, Milgrom83c}, see
\cite{FamaeyMcGaugh12} for a review) and in section~\ref{sec:applications} we
discuss the role of RG in other cosmic systems.
Finally, in section~\ref{sec:conclusions} we summarize the main points and
report possible directions for future work.

\section{The basics of \textit{refracted gravity}}
  \label{sec:indications}

\subsection{Elements from electrodynamics}

In the interior of a dielectric medium with no free charges, the electric field
$\mathbf E$ obeys the equation:
\begin{equation}
  \nabla \cdot (\varepsilon\mathbf E) = 0\,,
  \label{eq:electricpoisson}
\end{equation}
where $\varepsilon \geq 1$ is the electric permittivity.
In vacuum, $\varepsilon = 1$ and Poisson's law is recovered.
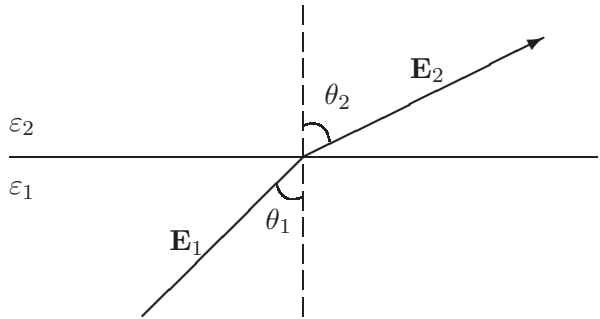
\begin{figure}
  \begin{center}
    \begin{picture}(240,120)
      \put(0,60){\line(1,0){220}}
      \multiput(110,0)(0,10){12}{\line(0,1){7}}
      \put(0,46){$\varepsilon_1$}
      \put(0,70){$\varepsilon_2$}
      \qbezier(100,50)(102,42)(110,45)
      \put(96,33){$\theta_1$}
      \qbezier(110,72)(118,75)(120,66)
      \put(118,80){$\theta_2$}
      \thicklines
      \put(50,0){\line(1,1){60}}
      \put(110,60){\vector(2,1){90}}
      \put(60,25){$\mathbf E_1$}
      \put(150,90){$\mathbf E_2$}
    \end{picture}
  \end{center}
  \caption{
    Refraction of an electric field line that threads two media of different
    permittivities, $\varepsilon_1 < \varepsilon_2$.}
  \label{fig:refraction}
\end{figure}
Equation~(\ref{eq:electricpoisson}) implies that, whenever a field line crosses
a surface that separates two media with different permittivities, it is
refracted according to:
\begin{equation}
  \frac{\tan\theta_1}{\tan\theta_2} = \frac{[E_2]_\bot}{[E_1]_\bot}
    = \frac{\varepsilon_1}{\varepsilon_2}\,,
  \label{eq:refraction}
\end{equation}
where the symbol ``$\bot$'' denotes the component perpendicular to the
interface, see figure~\ref{fig:refraction}.
Note that the magnitude of the electric field is reduced due to polarization
effects; specifically, the component parallel to the interface remains unchanged
whereas the perpendicular component decreases.

Motivated by electrodynamics, we explore a similar framework for gravity.
By introducing a ``gravitational permittivity'', a function monotonically
dependent on the mass density, we investigate the implications of the refracted
gravitational field in galactic systems.

\subsection{Formulation of \textit{refracted gravity}}

In analogy to electric fields in matter, in RG the gravitational field obeys the
following Poisson equation:
\begin{equation}
  \nabla\cdot(\epsilon\nabla\Phi) = 4\pi G\rho\,,
  \label{eq:poisson}
\end{equation}
where $\rho$ is the ordinary matter density, $\Phi$ the gravitational potential,
and $\epsilon = \epsilon(\rho)$ the gravitational permittivity that depends on
the local density.
An equation only formally similar to Eq.~(\ref{eq:poisson}) was proposed for
MOND in ref.~\cite{BekensteinMilgrom84}.
However, in that case the function $\epsilon$ depends on the field, i.e.
$\epsilon_\mathrm{MOND} = \epsilon(\nabla\Phi)$, rather than on the source as we
propose here.
This difference is clearly fundamental; we compare the two approaches in detail
in section~\ref{sec:mond}.

Contrary to electromagnetism, the permittivity of Eq.~(\ref{eq:poisson}) takes
the values $0 < \epsilon_0 \leq \epsilon(\rho) \leq 1$, where
$\epsilon_0 = \epsilon(0)$ is the permittivity of vacuum.
Specifically, it is approximately unity within matter and has a smaller but
constant value in vacuum, $\epsilon(0) = \epsilon_0 < 1$.

RG is thus based on two assumptions:
\begin{itemize}
\item{(1)}
The field equation for the gravitational potential is given by
Eq.~(\ref{eq:poisson});
\item{(2)}
$\epsilon = \epsilon(\rho)$ is a monotonically increasing function of $\rho$,
with $0 < \epsilon_0 \leq \epsilon(\rho) \leq 1$ and
$\epsilon_0 = \epsilon(0) < 1$.
\end{itemize}

The deviations from Newtonian gravity become apparent by considering the inner
product of Eq.~(\ref{eq:poisson}):
\begin{equation}
  \nabla\cdot(\epsilon\nabla\Phi)
    = \frac{\partial\epsilon}{\partial\rho}\nabla\rho\cdot\nabla\Phi
    + \epsilon\nabla^2\Phi\,.
  \label{eq:product}
\end{equation}
The two relevant modifications to the standard Poisson equation is the
additional term $(\partial\epsilon/\partial\rho)\nabla\rho\cdot\nabla\Phi$ and
the factor $\epsilon$ in front of the Laplacian.
If we broadly classify all astronomical objects -- from stellar scales up to
galaxy clusters -- into spherical- and flat-shaped objects, we see that large
deviations from standard gravity appear in flat structures due to the
redirection of the field lines (first term of the right-hand side in
Eq.~\ref{eq:product}), and in spherical systems due to the value of $\epsilon$
which effectively amplifies the field at large distances (second term of the
right-hand side in Eq.~\ref{eq:product}).

\subsubsection{Assumptions for the permittivity}

The mass densities in cosmic structures differ by orders of magnitudes from
planetary systems to galaxies, galaxy clusters, and large-scale structures.
In contrast, the gravitational permittivity is limited to
$\epsilon\in[\epsilon_0,\,1]$.
We may thus expect that a proper approximation for the permittivity is
\begin{equation}
  \epsilon(\rho) \simeq \left\{
  \begin{array}{ll}
    1 & \mathrm{for}\ \ \rho \gg \rho_\mathrm{c} \\
    \epsilon_0 & \mathrm{for}\ \ \rho \ll \rho_\mathrm{c}
  \end{array}
  \right.\,,
  \label{eq:epsilon}
\end{equation}
where $\rho_\mathrm{c}$ is some critical density.

More generally, we note that the permittivity $\epsilon(\rho)$ could in
principle depend on other local variables that characterize the gravitational
sources, such as the energy density or the entropy.
As long as any such quantity exhibits contours similar to $\rho$, then the
corresponding phenomenology would be in agreement with all the observational
tests discussed in sections~\ref{sec:spirals} and \ref{sec:rgpoisson} below.
Here, we only consider $\rho$ because it is the simplest hypothesis.

Furthermore, by introducing the permittivity $\epsilon$ as a function of density
$\rho$, we unavoidably introduce a smoothing length $D$ over which we average
the density field.
We limit the present analysis to a proof of concept and postpone the
investigation of this crucial topic to future work; here, we simply speculate
that the scale of $D$ is expected to be of astronomical interest, for example
tens of astronomical units or larger, because the mass discrepancy problem
appears to be relevant on scales larger than the solar system.
A number of additional serious issues remain open, including a possible
incompatibility with special relativity due to the introduction of the vacuum
permittivity or a possible violation of the Strong Equivalence Principle.
The consequences of a variational approach applied to a possible RG Lagrangian
of the form
\begin{equation}
  L = \frac{\epsilon}{8\pi G}(\nabla\Phi)^2 + \rho\Phi
\end{equation}
should also be investigated.
However, in this paper we limit our study to assess whether this new idea
appears promising, at least phenomenologically.
If so, we will explore all these fundamental issues elsewhere.

\subsubsection{Flat systems}

Flat systems have a strong vertical density stratification and hence the value
of $\rho_\mathrm{c}$ is reached at a given height.
As explained in the following sections, Eq.~(\ref{eq:poisson}) suggests that in
such objects the field is redirected and, near the disk, it becomes almost
parallel to the density layers.

To estimate the effect of this reorientation of the field lines on the exerted
force, consider the following example.
The directions of the field lines originating from a point source are uniformly
distributed over a solid angle of $4\pi$\,sr.
Consequently, the conservation of the field flux implies that the field strength
drops proportionally to the inverse of the spherical surface, $4\pi R^2$.
Now consider a mechanism that redirects the field lines at large distances
parallelly to a specific plane.
In this case, the conservation of flux implies a power law for the field that is
proportional to the inverse of the circular circumference, namely
$\propto r^{-1}$.
As it is shown in sections~\ref{sec:chain} and \ref{sec:spirals}, this is the
main mechanism in RG that changes the radial profile of the acceleration within
a disk galaxy from $\propto r^{-2}$ close to its center to $\propto r^{-1}$
farther out.

\subsubsection{Spherical systems}
  \label{sec:spherical}

If in the central regions of a spherical system, such as a galaxy cluster, the
density is larger than the critical value, $\rho > \rho_\mathrm{c}$, then the
permittivity is unity, $\epsilon = 1$, and hence the field is given by the
standard Poisson equation:
\begin{equation}
  \nabla^2\Phi = 4\pi G\rho\,.
  \label{eq:poissoninterior}
\end{equation}
In this case no deviations from Newtonian gravity are expected in RG.
On the contrary, in the outer regions where the density eventually drops below
the critical value $\rho_\mathrm{c}$, the permittivity becomes
$\epsilon = \epsilon_0 < 1$.
Due to the spherical symmetry $\nabla\rho$ and $\nabla\Phi$ are parallel and
hence the field lines remain radial (see Eq.~\ref{eq:product}).
Nevertheless, the strength of the field deviates from the Newtonian law and
appears stronger:
\begin{equation}
  \nabla^2\Phi = 4\pi G\frac{\rho}{\epsilon_0}\,.
  \label{eq:poissonexterior}
\end{equation}
In principle, measurements of the dynamical properties of spherical systems --
namely stars and globular clusters that exhibit a Newtonian behavior, and
elliptical galaxies and galaxy clusters that show increasing departure from
Newtonian dynamics -- can be used to estimate the values of $\rho_\mathrm{c}$
and $\epsilon_0$.

\subsection{A simplified RG framework (SRG) for flat systems}
  \label{sec:framework}

In order to solve Eq.~(\ref{eq:poisson}), an explicit expression is required for
$\epsilon(\rho)$ with specific values for $\epsilon_0$ and $\rho_\mathrm{c}$.
We undertake this task in section~\ref{sec:rgpoisson}.
Here, we refrain from exploring special cases given by specific permittivity
expressions, and adopt a simplified but general approach to highlight the basic
properties of the theory.
For spherical systems we may employ Eq.~(\ref{eq:poissonexterior}) to infer the
phenomenology in regions where $\rho < \rho_\mathrm{c}$: it follows that the SRG
field is equivalent to the Newtonian gravitational field of an effective mass
distribution $\rho/\epsilon_0 > \rho$ (see also section~\ref{sec:dm}).
However, such an approach cannot apply to flat systems because it does not
account for the redirection of the field lines.
Instead, we may approximate Eq.~(\ref{eq:epsilon}) with a two-valued
permittivity $\epsilon \in \{0,\,1\}$ and assume that the transition is defined
by the threshold $\rho_\mathrm{c}$.
Even though we consider for simplicity $\epsilon_0 = 0$, we note that the
redirection of the field lines takes place for any value less than unity; the
smaller the value of the vacuum permittivity the stronger is the effect.
For $\epsilon_0 = 0$ we may estimate the gravitational field in the region
$\epsilon = 1$ by applying the following condition on the iso-surface of
$\rho_\mathrm{c}$:
\begin{equation}
  \nabla_\bot\Phi \simeq 0\,.
  \label{eq:bc}
\end{equation}

These approximations to Eq.~(\ref{eq:poisson}) are equivalent to solving the
standard Newtonian Poisson equation with the boundary condition of
Eq.~(\ref{eq:bc}) on the density contour
$\rho = \rho_\mathrm{c}$.\footnote{
The equivalence becomes evident when Eq.~(\ref{eq:poisson}) is associated to a
time-dependent equation of a similar form, such as the heat equation,
$\partial T/\partial t = \nabla\cdot(k\nabla T)$.
For a given time interval, an interface that separates two media with
$(T_1,\,k_1)$ and $(T_2,\,k_2)$ behaves as an insulator when $k_1 = 1$ and
$k_2 \to 0$, i.e. $[\nabla T_1]_\bot \simeq 0$.
Analogously, in the limit $k_2 \to 0$ the characteristic time scale of heat
transfer in region 2 becomes infinite, and thus the steady-state temperature
distribution in region 1 can be effectively described by a Poisson equation with
a zero-flux boundary condition.}
Due to the vertical density stratification of flat objects, such as disk
galaxies, we may approximate the iso-surface of $\rho_\mathrm{c}$ as two
parallel planes above and below the disk.
In turn, the boundary condition at that height, e.g. $z = \pm h$, provides the
mechanism to redirect all the field flux within the volume defined by
$-h \leq z \leq h$.
This effect entails from Eq.~(\ref{eq:bc}), which represents a reflective
boundary condition: it reverses the sign of the perpendicular component of the
gradient whereas it keeps the parallel component unchanged.
In section~\ref{sec:rgpoisson} we show that this approach is indeed a valid
approximation by comparing it with the numerical solution of the full RG field
equation.

SRG essentially introduces the parameter $h$ to determine the volume surrounding
flat systems within which the gravitational flux is redirected.
In the central regions, i.e. on scales much smaller than $h$, field-redirection
effects are negligible.
Conversely, strong deviations from the Newtonian regime are expected in the
outskirts of such objects.
With this general approach we may perform explicit calculations to describe the
approximate topology of the field given by Eq.~(\ref{eq:poisson}) for any
expression of $\epsilon$ that obeys Eq.~(\ref{eq:epsilon}).

\subsubsection{Quantifying the deviations from the inverse square law}
  \label{sec:chain}

In order to quantify the effects of SRG, we assume that the disk galaxy is a
point mass $m$ in between a bounding surface of two infinite parallel planes
whose separation is equal to $2h$.
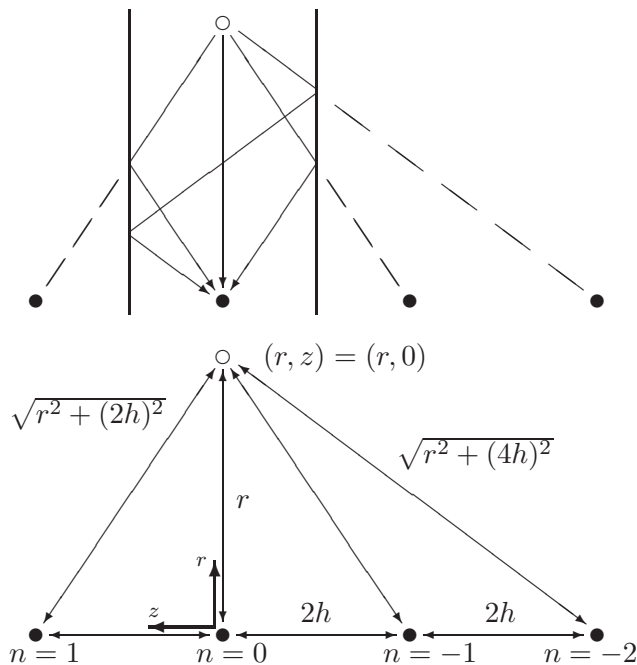
\begin{figure}
  \begin{center}
    \begin{picture}(240,125)
      \put(10,10){\circle*{5}}
      \put(80,10){\circle*{5}}
      \put(150,10){\circle*{5}}
      \put(220,10){\circle*{5}}
      \put(80,115){\circle{5}}
      \put(80,60){\line(0,1){50}}
      \put(80,60){\vector(0,-1){45}}
      \put(45,62){\line(2,3){32}}
      \put(45,62){\vector(2,-3){32}}
      \put(115,62){\line(-2,3){32}}
      \put(115,62){\vector(-2,-3){32}}
      \put(115,90){\line(-4,3){29}}
      \multiput(122,84)(20,-15){5}{\line(4,-3){13}}
      \put(45,5){\line(0,1){115}}
      \put(115,5){\line(0,1){115}}
      \multiput(41,56)(-10,-15){3}{\line(-2,-3){7}}
      \multiput(119,56)(10,-15){3}{\line(2,-3){7}}
      \put(45,36){\line(4,3){70}}
      \put(45,35){\vector(4,-3){30}}
    \end{picture}
    \begin{picture}(240,125)
      \put(10,10){\circle*{5}}
      \put(0,0){$n = 1$}
      \put(80,10){\circle*{5}}
      \put(70,0){$n = 0$}
      \put(150,10){\circle*{5}}
      \put(140,0){$n = -1$}
      \put(220,10){\circle*{5}}
      \put(200,0){$n = -2$}
      \put(45,10){\vector(1,0){30}}
      \put(45,10){\vector(-1,0){30}}
      \put(115,10){\vector(1,0){30}}
      \put(115,10){\vector(-1,0){30}}
      \put(109,15){$2h$}
      \put(185,10){\vector(1,0){30}}
      \put(185,10){\vector(-1,0){30}}
      \put(178,15){$2h$}
      \put(80,115){\circle{5}}
      \put(95,112){$(r, z) = (r, 0)$}
      \put(80,60){\vector(0,1){50}}
      \put(80,60){\vector(0,-1){45}}
      \put(85,58){$r$}
      \put(46,64){\vector(2,3){31}}
      \put(46,64){\vector(-2,-3){33}}
      \put(0,90){$\sqrt{r^2 + (2h)^2}$}
      \put(114,64){\vector(-2,3){31}}
      \put(114,64){\vector(2,-3){33}}
      \put(151,63){\vector(-4,3){65}}
      \put(151,63){\vector(4,-3){65}}
      \put(145,75){$\sqrt{r^2 + (4h)^2}$}
      \thicklines
      \put(77,13){\vector(-1,0){25}}
      \put(77,13){\vector(0,1){25}}
      \thinlines
      \put(52,13){$^z$}
      \put(70,32){$^r$}
    \end{picture}
  \end{center}
  \caption{
    Top panel: a point mass between reflective boundaries (second solid circle
    from the left) is equivalent to a chain of equally spaced and aligned point
    masses.
    The open circle represents a test particle.
    Bottom panel: the configuration from which Eq.~(\ref{eq:fr}) is derived.}
  \label{fig:chain}
\end{figure}
As shown in the top panel of figure~\ref{fig:chain}, the reflective property of
the boundary condition (Eq.~\ref{eq:bc}) makes the system equivalent to a
chain of infinite, aligned, and equally-spaced point masses in the Newtonian
framework (bottom panel).
By setting an integer label $n$ for each node of mass $m$, with $n = 0$
corresponding to the reference mass, the distance between the nodes $n = 0$ and
$n = k$ is equal to $2kh$.

In cylindrical coordinates, where the $z$ axis is taken along the chain of
masses, it is straightforward to derive the force at distance $r$ from the
reference node:
\begin{equation}
  F(r,0) = -\sum_{n = -\infty}^\infty\left[\frac{Gm}{r^2 + (2nh)^2}
    \frac{r}{\sqrt{r^2 + (2nh)^2}}\right] = -\frac{Gm}{r^2}
    - 2\sum_{n = 1}^\infty\frac{Gmr}{\left[r^2 + (2nh)^2\right]^{3/2}}\,.
  \label{eq:fr}
\end{equation}
For $r \ll h$ the first term dominates and the inverse square law is recovered,
i.e. $F \simeq -Gm/r^2$.
To the first order correction, we have 
\begin{equation}
  F(r) \simeq -\frac{Gm}{r^2} - \frac{Gmr}{4h^3}\sum_{n=1}^\infty\frac{1}{n^3}
    = -\frac{Gm}{r^2} - \frac{Gmr}{4h^3}\zeta(3)\,,
  \label{eq:frapproximation}
\end{equation}
where $\zeta(q) = \sum_{n = 1}^\infty n^{-q}$ is the Riemann zeta function, and
$\zeta(3) \simeq 1.2$. 
In the other limit, $r \gg h$, the force becomes $F \simeq -Gm/hr$.
In fact the Newtonian gravitational field of a continuous massive string with
linear density $m/2h$ exactly is $F =-Gm/hr$, an elementary application of the
Gauss theorem; in RG, the effect of representing the massive point source as a
continuous massive string only appears evident at large distances $r \gg h$.

\begin{figure}
  \centering
  \includegraphics[width=0.65\textwidth]{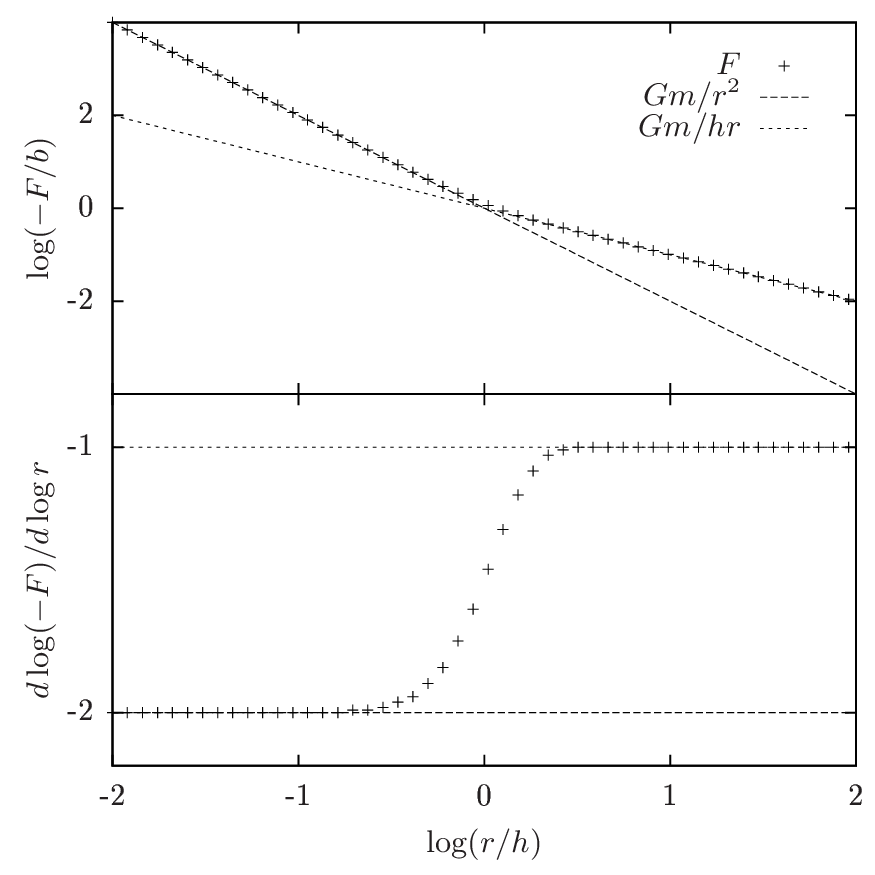}
  \caption{
    Top panel: radial profile of the force generated by a point mass in between
    two reflective parallel planes (Eq.~\ref{eq:fr}).
    Bottom panel: the effective radial exponent of Eq.~(\ref{eq:fr}).
    For small radii, $r \ll h$, $F$ is proportional to $r^{-2}$, while for large
    ones, $r \gg h$, the power law becomes $F \propto r^{-1}$.}
  \label{fig:force}
\end{figure}
In order to visualize the profile of Eq.~(\ref{eq:fr}), the top panel of
figure~\ref{fig:force} plots its values normalized to the acceleration
\begin{equation}
  b = \frac{Gm}{h^2}
  \label{eq:b}
\end{equation}
as a function of the distance normalized to $h$.
The bottom panel shows the corresponding index of the power law at each
location.
The change in the radial dependence occurs at the region where $r$ is of the
order of $h$, around which the force switches from $F \propto r^{-2}$ to
$F \propto r^{-1}$.

\subsubsection{Basic implications}
  \label{sec:dynamics}

Focusing on the plane $z = 0$, the Newtonian force is $N = -Gm/r^2$ and the
orbital velocity at distance $r$ is $v_N = \sqrt{-Nr}$.
For $r \ll h$ the Keplerian rotation is recovered, i.e. $F \simeq N$ and thus
$v_F \simeq v_N = \sqrt{Gm/r}$.
However, for $r \gg h$ the orbital velocity in the presence of $F$ becomes
\begin{equation}
  v_F = \sqrt{-Fr} \simeq \sqrt{\frac{Gm}{h}} = v_\mathrm{f}\,.
  \label{eq:vf}
\end{equation}
Notably, this asymptotic value does not depend on $r$: the velocity
$v_\mathrm{f}$ has a flat rotation profile which is denoted with the subscript
``f''.

Moreover, the elimination of $h$ from Eq.~(\ref{eq:vf}) (using Eq.~\ref{eq:b})
gives:
\begin{equation}
  v_\mathrm{f}^4 = Gb\,m\,.
  \label{eq:tf}
\end{equation}
By assuming a constant mass-to-light ratio, Eq.~(\ref{eq:tf}) reproduces the
slope of the empirical Tully-Fisher relation \cite{TullyFisher77}.
To be also consistent with the observed normalization of the Tully-Fisher
relation, $b$ must be the constant
\begin{equation}
  b \simeq 1.2\cdot10^{-10}\,\mathrm{m}\,\mathrm{s}^{-2}\,,
\end{equation}
as shown in ref.~\cite{FamaeyMcGaugh12}.
The combination of Eqs.~(\ref{eq:vf}) and (\ref{eq:tf}) links the location of
the parallel plane boundaries to the total mass, i.e. $h = \sqrt{Gm/b}$.
In the case of disk galaxies the parameter $h$ can be related to the scale
height of the disk, as it is argued in section \ref{sec:assumptions} below.

Furthermore, the ratio of the squared velocities of the RG and Newtonian fields,
termed mass discrepancy \cite{McGaugh04}, is equal to $(v_F/v_N)^2 = F/N$.
For $r$ smaller than $h$ the fraction is unity, whereas in the other limit,
$r \gg h$, it becomes
\begin{equation}
  \frac{v_F^2}{v_N^2} = \frac{F}{N} \simeq \frac{Gm/hr}{Gm/r^2} = \frac{r}{h}\,.
  \label{eq:ratio-r}
\end{equation}
In this region it is straightforward to rewrite the above relation as a function
of the forces, $F$ or $N$:
\begin{equation}
  \frac{v_F^2}{v_N^2} \simeq \frac{b}{-F}\,,
  \label{eq:ratio-f}
\end{equation}
\begin{equation}
  \frac{v_F^2}{v_N^2} \simeq \sqrt{\frac{b}{-N}}\,.
  \label{eq:ratio-n}
\end{equation}
These expressions match well the observed profiles \cite{FamaeyMcGaugh12,
McGaugh04}, see section~\ref{sec:massdisc}.

\section{Application of SRG to disk galaxies}
  \label{sec:spirals}

The stellar distribution of a disk galaxy can be roughly approximated as a thin
disk of surface density $\sigma(r) = \sigma_0\exp{(-r/r_0)}$, where $\sigma_0$
and $r_0$ are constants \cite{Freeman70}.
For the sake of simplicity the bulge and the gas components are neglected in the
following estimates.
The Newtonian gravitational potential in cylindrical coordinates, with the
origin located at the center of the disk, is
\begin{equation}
  \Phi_N(r,z) = -\int_0^{2\pi}\!\!\!\int_0^{\infty}\!\!\!
    \frac{G\sigma_0e^{-r'/r_0}\,r'\,\mathrm{d}r'\,\mathrm{d}\phi}
    {\sqrt{r'^2 - 2rr'\cos\phi + r^2 + z^2}}\,,
  \label{eq:phinw}
\end{equation}
and the total enclosed mass is 
\begin{equation}
  m = \int_0^{2\pi}\!\!\!\int_0^\infty
    \sigma_0e^{-r'/r_0}\,r'\,\mathrm{d}r'\,\mathrm{d}\phi
    = 2\pi\sigma_0r_0^2\,.
  \label{eq:sigma}
\end{equation}

Following the SRG approach, we consider the iso-surface of $\rho_\mathrm{c}$ to
consist of two parallel planes -- located at $\pm h$ above and below the disk --
and we assume that $\nabla_\bot\Phi = 0$ holds on this boundary.
The system is equivalent to a sequence of infinite and parallelly aligned disks
of surface density $\sigma(r)$, similarly to figure~\ref{fig:chain}.
In this context, the potential is 
\begin{equation}
  \Phi_F(r,z) =
    -\sum_{n = -\infty}^{\infty}\int_0^{2\pi}\!\!\!\int_0^{\infty}\!\!\!
    \frac{G\sigma_0e^{-r'/r_0}\,r'\,\mathrm{d}r'\,\mathrm{d}\phi}
    {\sqrt{r'^2 - 2rr'\cos\phi + r^2 + (z + 2nh)^2}}\,,
  \label{eq:phirg}
\end{equation}
with $h = \sqrt{Gm/b}$.
In the following sections we examine the observational implications of the force
$F = -\nabla\Phi_F$ on the disk midplane, $z = 0$.

\begin{table}
  \centering
  \begin{tabular}{lcccc}
    \hline
    Name & $\sigma_0$\,[$M_\odot\,\mathrm{pc}^{-2}$] & $r_0$\,[$\mathrm{kpc}$] &
      $\quad m$\,[$M_\odot$] $\quad$ & $h$\,[$\mathrm{kpc}$] \\
    \hline
    G0 & $2.4\cdot10^2$ & $3.3$ & $1.7\cdot10^{10}$ & $4.4$ \\
    G1 & $9.6\cdot10^2$ & $1.5$ & $1.4\cdot10^{10}$ & $4.0$ \\
    G2 & $2.4\cdot10^3$ & $3.3$ & $1.7\cdot10^{11}$ & $14$  \\
    G3 & $9.6\cdot10^2$ & $7.6$ & $3.5\cdot10^{11}$ & $20$  \\
    G4 & $2.4\cdot10^2$ & $10$  & $1.5\cdot10^{11}$ & $13$  \\
    G5 & $5.7\cdot10^1$ & $7.6$ & $2.1\cdot10^{10}$ & $5.0$ \\
    G6 & $2.4\cdot10^1$ & $3.3$ & $1.7\cdot10^{9}$  & $1.4$ \\
    G7 & $5.7\cdot10^1$ & $1.5$ & $8.6\cdot10^{8}$  & $1.0$ \\
    G8 & $2.4\cdot10^2$ & $1.0$ & $1.5\cdot10^{9}$  & $1.3$ \\
    \hline
    M1 & $4.8\cdot10^3$ & $5.0$ & $7.7\cdot10^{11}$ & $30$  \\
    M2 & $1.5\cdot10^3$ & $9.0$ & $7.8\cdot10^{11}$ & $30$  \\
    M3 & $4.8\cdot10^2$ & $16$  & $7.8\cdot10^{11}$ & $30$  \\
    \hline
  \end{tabular} 
  \caption{
    Disk galaxy models with typical values for $\sigma_0$, $r_0$, $m$, and $h$.
    They are shown with open circles (G0--G8) and open squares (M1--M3) in
    figure~\ref{fig:objects}.}
  \label{tab:spirals}
\end{table}
\begin{figure}
  \centering
  \includegraphics[width=0.65\textwidth]{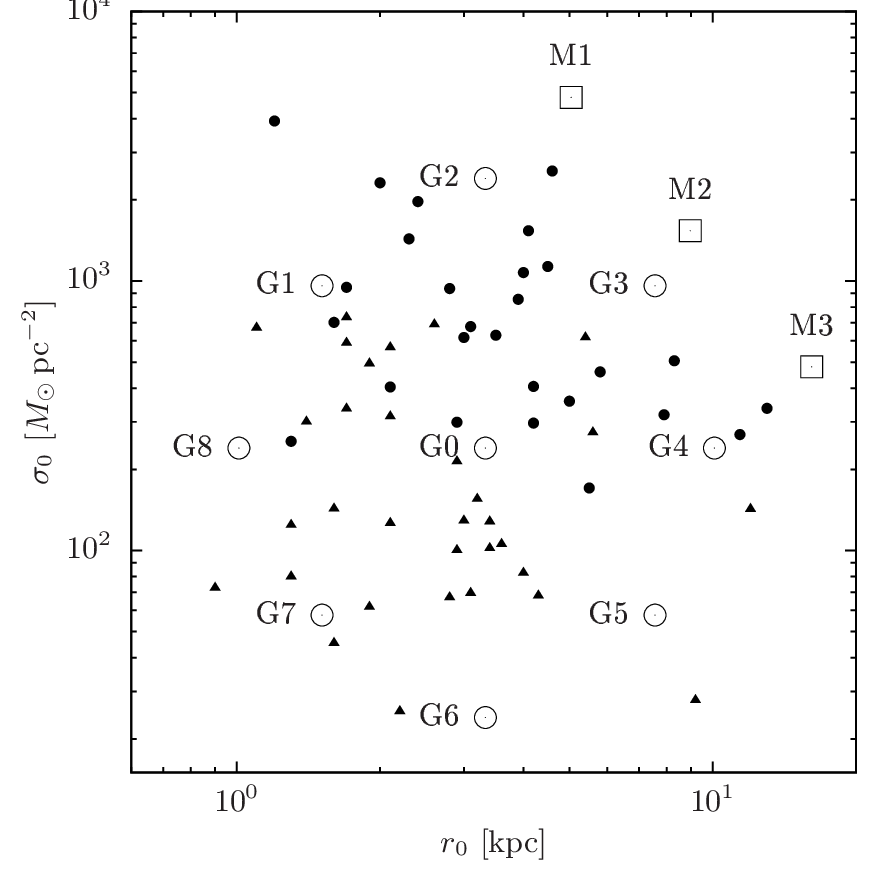}
  \caption{
    Typical scale lengths $r_0$ and surface densities $\sigma_0$ for
    star-dominated disk galaxies (dots) as well as disk galaxies whose gas
    contribution is larger than $20\%$ (triangles) \cite{McGaugh05b}.
    Open circles and squares denote the models listed in Table~\ref{tab:spirals}
    and are selected such that they cover the full range of the observed
    parameters.}
  \label{fig:objects}
\end{figure}
Table~\ref{tab:spirals} lists a selection of disk galaxy parameters that cover
the typical values of such objects.
The models G0--G8 are depicted with circles in figure~\ref{fig:objects} and are
shown along with observational data (dots and triangles) from
ref.~\cite{McGaugh05b}.
The $\sigma_0$ values of the observed galaxies have been derived from
Eq.~(\ref{eq:sigma}) using the total baryonic mass of the galaxy, i.e.
accounting for both the stars and the gas.
A dot is used if the amount of the stellar mass is larger than $80\%$ of the
total mass and a triangle otherwise.
Since the gas distribution does not follow the same exponential law as $\sigma$,
models G0--G4 can be considered more realistic than G5--G8, which cover the
region of gas-rich galaxies in the plane $r_0$--$\sigma_0$.
Nevertheless, in order to investigate the various aspects of RG,
Eq.~(\ref{eq:phirg}) is used to explore the whole range shown in
figure~\ref{fig:objects}.
Table~\ref{tab:spirals} also includes the models M1--M3 which have a relatively
higher mass and are shown with open squares in figure~\ref{fig:objects}.

\subsection{Rotation curves}
  \label{sec:rotcurves}

\begin{figure}
  \centering
  \includegraphics[width=0.65\textwidth]{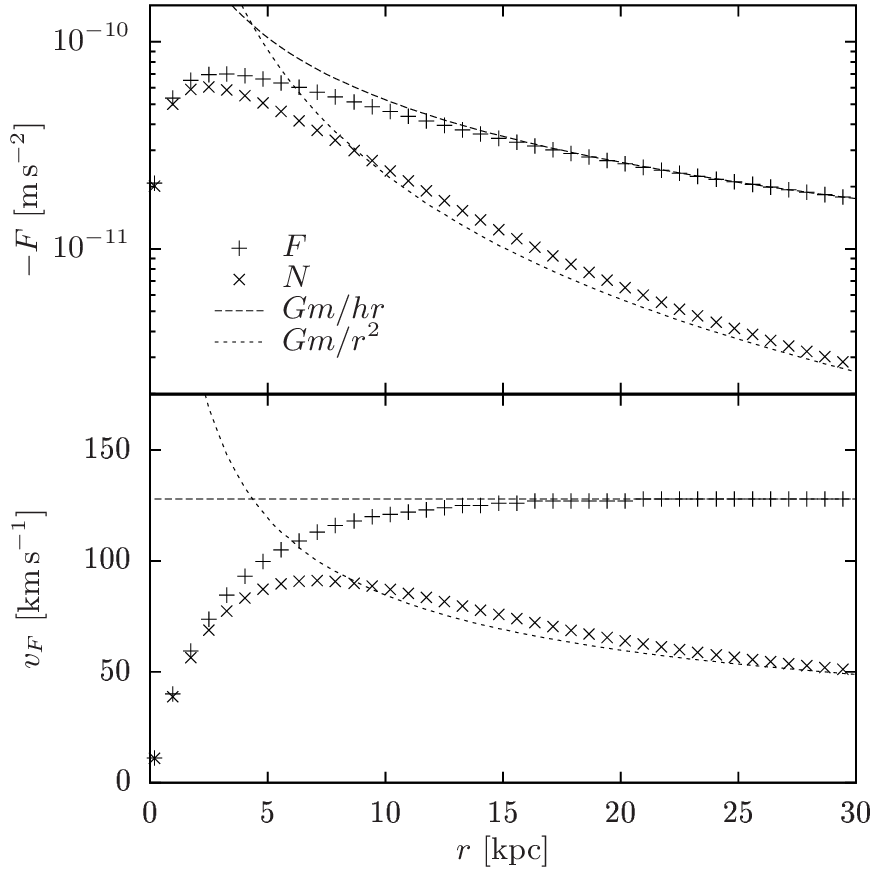}
  \caption{
    Top panel: radial forces for the disk galaxy G0 as given by the SRG (pluses)
    and Newtonian fields (crosses).
    The dashed lines assume that the total galactic mass is concentrated in the
    center.
    Bottom panel: orbital velocity profiles derived from the corresponding force
    fields.
    The flatness of the rotation curve becomes apparent beyond a radius
    comparable to a few $h$.}
  \label{fig:galaxy}
\end{figure}
The Newtonian and SRG forces on the midplane of a disk galaxy can be directly
computed from Eqs.~(\ref{eq:phinw}) and (\ref{eq:phirg}), respectively.
Their values as a function of the radial distance are shown in the top panel of
figure~\ref{fig:galaxy} for model G0.
In agreement with the results of section~\ref{sec:chain}, deviations appear at
$r\sim h=4.4$\,kpc; at this location $F$ becomes larger than $N$, an effect that
increases with radius.
The dashed lines illustrate the accelerations $-Gm/r^2$ and $-Gm/rh$, assuming
that the total galactic mass is concentrated in a point.
Since most of the mass is enclosed within a few scale lengths, $F$ and $N$
converge to the corresponding asymptotic curves beyond $\sim 5r_0 = 16.5$\,kpc.

\begin{figure}
  \centering
  \includegraphics[width=0.65\textwidth]{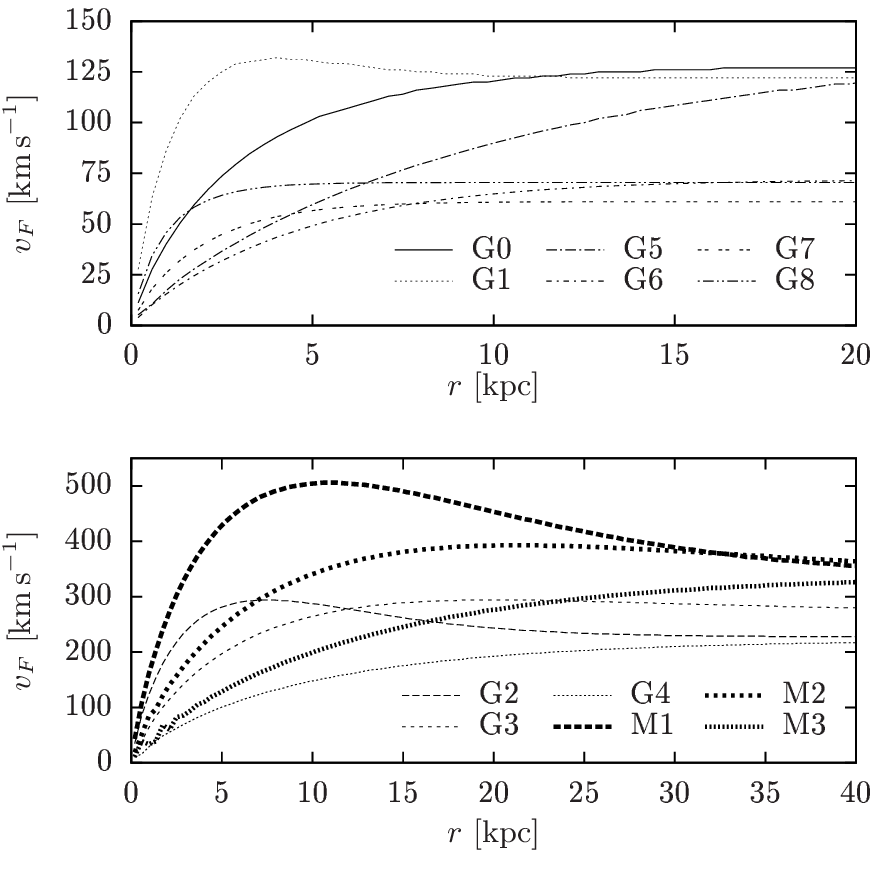}
  \caption{
    Rotation curves for the galaxy models listed in Table~\ref{tab:spirals}.
    All of them reach a flat profile, each one at a different radius.
    High-surface-density systems exhibit a peak after which the values converge
    to the asymptotic limit.
    Low-surface-density systems show a slowly-increasing velocity profile.}
  \label{fig:rotcurves}
\end{figure}
The bottom panel of figure~\ref{fig:galaxy} displays the orbital velocities that
correspond to the Newtonian and SRG forces, respectively.
At large distances $v_F$ exhibits a flat profile which is the typical feature of
most galactic rotation curves.
The rotation curves for the rest of the models of Table~\ref{tab:spirals} are
shown in figure~\ref{fig:rotcurves}.
Generally, the speed increases with radius reaching a constant value at large
distances.
The profiles of the high-surface-density models, G1--G3 and M1--M2, exhibit a
peak after which the curves drop slightly and converge to their asymptotic
limit.
Conversely, models with low surface density show a rotation curve that is
monotonically increasing.
These features, and their dependence on the surface brightness, are in agreement
with the phenomenology of spiral galaxies \cite{Persic+96}.

\begin{figure}
  \centering
  \includegraphics[width=0.65\textwidth]{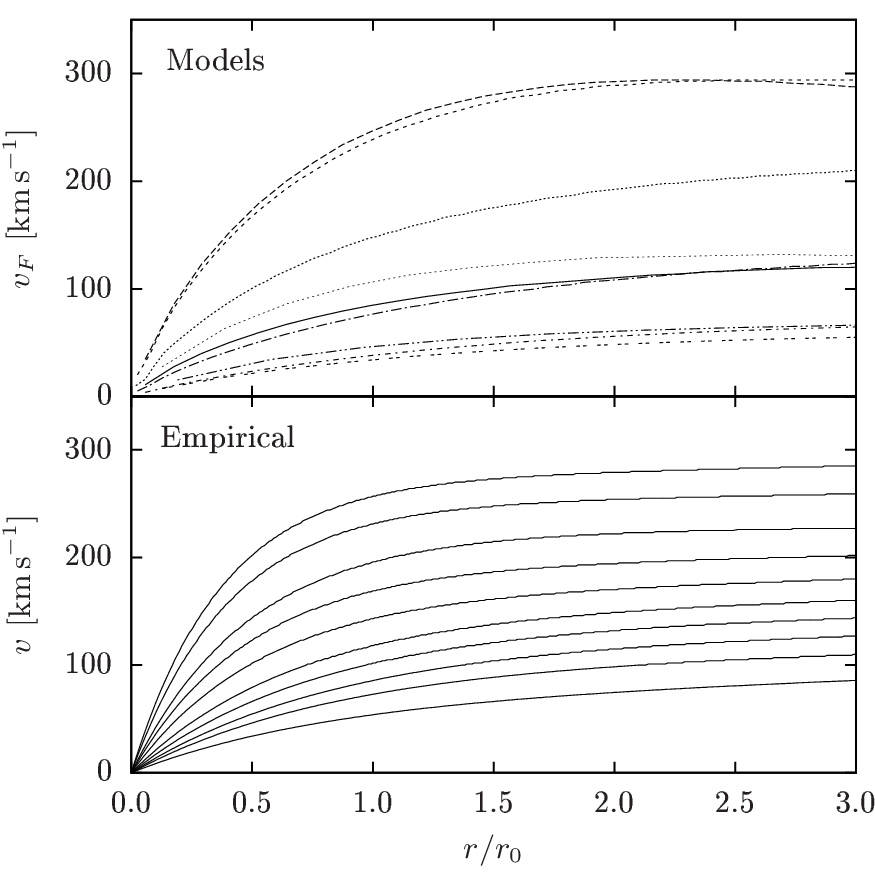}
  \caption{
    Rotation curves of models G0--G8 (top panel) as a function of the normalized
    radius $r/r_0$ (line types are as listed in figure~\ref{fig:rotcurves}).
    The template orbital velocity profiles presented in ref.~\cite{Catinella+06}
    are displayed in the bottom panel.
    Both panels show a slowly-increasing curve for the low-surface-brightness
    galaxies.
    Similarities also exist for the high-surface-brightness systems, where a
    steep gradient at the inner regions is followed by a flat profile in the
    outskirts.}
  \label{fig:template}
\end{figure}
Moreover, the plethora of available observational data has led to empirical
expressions that parametrize and fit the rotation curves of disk galaxies
\cite{Catinella+06}.
Comparison with such analytical formulas allows to test SRG in a general manner,
avoiding the peculiarities that specific objects might have.
The top panel of figure~\ref{fig:template} plots the orbital velocities of
models G0--G8 as a function of the corresponding normalized radius $r/r_0$.
The bottom panel shows the template rotation curves reported in
ref.~\cite{Catinella+06} that have been derived by fitting the data of
$\sim$$2200$ galaxies.
A visual inspection suggests that the radial dependence of $v_F$ follows these
inferred expressions.
The lower-mass galaxy models G6--G8 demonstrate a slowly increasing rotation
curve, in agreement with the empirical rule.
Similarly, the curves of the higher mass models G2--G4 reflect the initial steep
increase as well as the flat profile that follows.
Note that the presence of a bulge, that we neglect here, is expected to provide
a stronger gradient in the inner regions.
In section \ref{sec:NGCrotcurves} below we show two specific examples of
rotation curves of a high- and a low-surface-brightness galaxy, where we solve
the full RG field equation rather than just approximating the phenomenology with
SRG.

\begin{figure}
  \centering
  \includegraphics[width=0.65\textwidth]{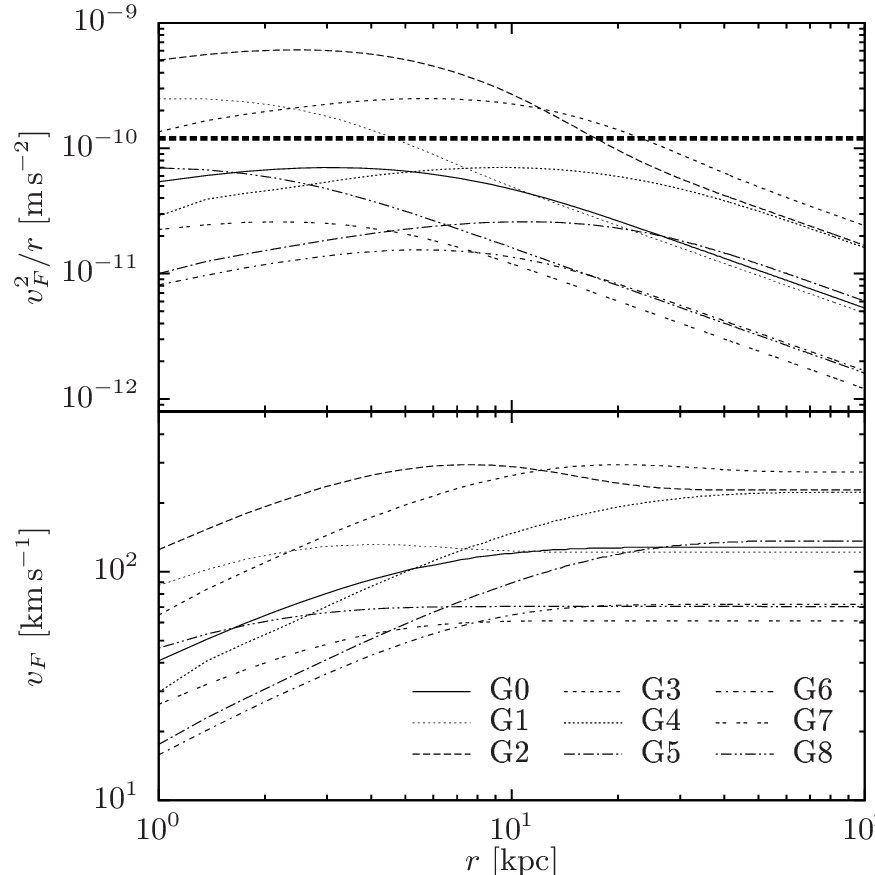}
  \caption{
    Centripetal acceleration (top panel) and orbital velocity (bottom panel) as
    a function of radius for models G0--G8.
    The thick horizontal dashed line denotes the value of $b$.}
  \label{fig:rfvlog}
\end{figure}
Additional evidence comes from the centripetal acceleration and orbital velocity
plotted in logarithmic scales in the top and bottom panels of
figure~\ref{fig:rfvlog} for the models G0--G8.
The curves closely resemble the big collection of observations illustrated in
ref.~\cite{FamaeyMcGaugh12}.
The inner regions of the high-surface-density models G2--G4 exhibit a Newtonian
behavior because $r < h$ there.
On the contrary, due to their small value of $h$, the low-surface-density models
are almost entirely in the RG regime with strong deviations throughout the disk.

\subsection{The baryonic Tully-Fisher relation}
  \label{sec:tf}

\begin{figure}
  \centering
  \includegraphics[width=0.65\textwidth]{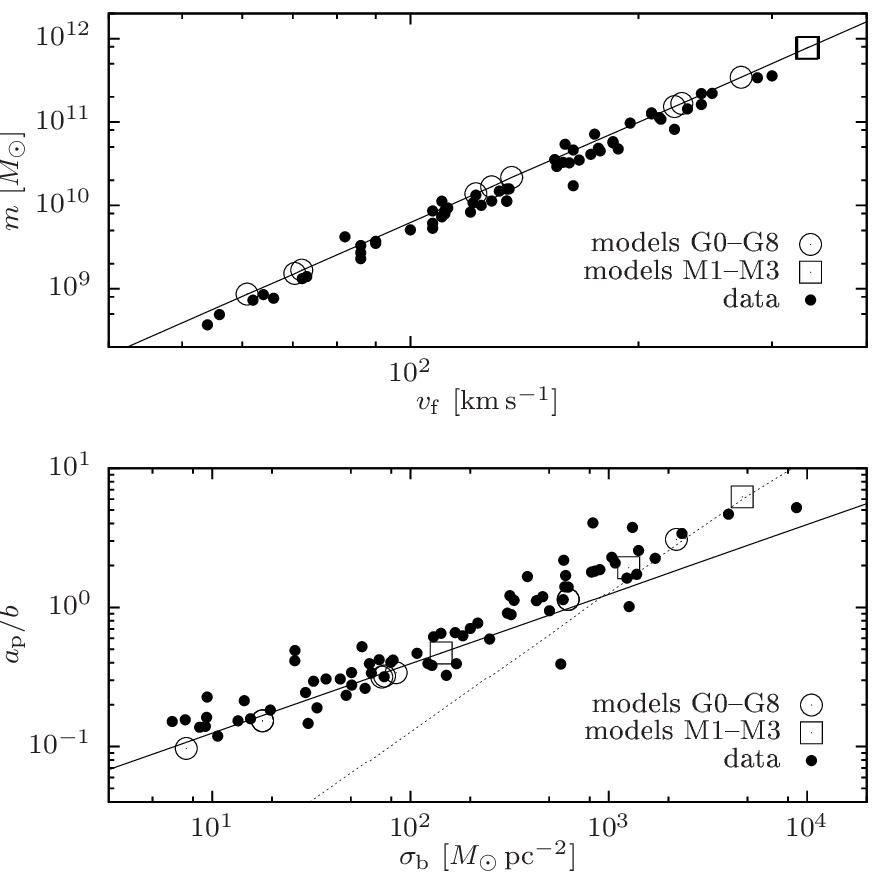}
  \caption{
    Top panel: the baryonic Tully-Fisher relation \cite{TullyFisher77}.
    Eq.~(\ref{eq:tf}) is plotted with a solid line, the models of
    Table~\ref{tab:spirals} with open circles and squares, and the observational
    data with dots \cite{McGaugh05b}.
    Bottom panel: dynamical acceleration versus characteristic baryonic surface
    density.
    The Newtonian expectation (dotted line) suggests
    $a_\mathrm{p} \propto \sigma_\mathrm{b}$.
    However, most of the models of Table~\ref{tab:spirals} (open circles and
    squares) as well as the observational data (dots) \cite{McGaugh10a} appear
    to follow the proportionality $a_\mathrm{p} \propto \sigma_\mathrm{b}^{1/2}$
    (solid line).
    We have multiplied the data points in the bottom panel by the factor $1.5$
    along the horizontal axis.
    This is equivalent to assuming a different stellar mass estimator and it
    does not affect the slope, see the discussion in ref.~\cite{McGaugh10a}.}
  \label{fig:tf}
\end{figure}
In the top panel of figure~\ref{fig:tf} we plot the baryonic galactic mass $m$
as a function of the asymptotic value of the rotation curve $v_\mathrm{f}$ for a
collection of observed galaxies (dots) \cite{McGaugh05b}.
The models of Table~\ref{tab:spirals} (circles and squares) exactly follow the
baryonic Tully-Fisher relation (solid line, Eq.~\ref{eq:tf}), as anticipated in
section~\ref{sec:dynamics}.

Furthermore, we may define the dynamical acceleration as
$a_\mathrm{p} = v_\mathrm{p}^2/r_\mathrm{p}$ and the characteristic baryonic
surface density as $\sigma_\mathrm{b} = 3m/4r_\mathrm{p}^2$, with $v_\mathrm{p}$ 
the velocity at radius $r_\mathrm{p}$ where the sum of the mass in stars and gas
make the maximum contribution to the rotation velocity (see
ref.~\cite{McGaugh05a}).
Observationally, $a_\mathrm{p}$ and $\sigma_\mathrm{b}$ are closely related as
shown by the data points \cite{McGaugh10a} in the bottom panel of
figure~\ref{fig:tf}.
Should the Newtonian framework be valid, we would have
$a_\mathrm{p} = G\sigma_\mathrm{b}$, and the galaxies would follow the dotted
line.
However, since Eq.~(\ref{eq:tf}) gives
$m = v_\mathrm{f}^4/Gb \sim v_\mathrm{p}^4/Gb$, SRG suggests
$a_\mathrm{p} = (4Gb/3)^{1/2}\sigma_\mathrm{b}^{1/2}$ (solid line), a
proportionality that is supported by observations (see also
ref.~\cite{FamaeyMcGaugh12}).
Most of the models in Table~\ref{tab:spirals} follow this curve (open circles
and squares), with the exception of the high-surface-density models whose
rotation velocity peaks in the Newtonian regime and thus they lie along the
dotted line.

\subsection{Mass discrepancies}
  \label{sec:massdisc}

\begin{figure}
  \centering
  \includegraphics[width=0.65\textwidth]{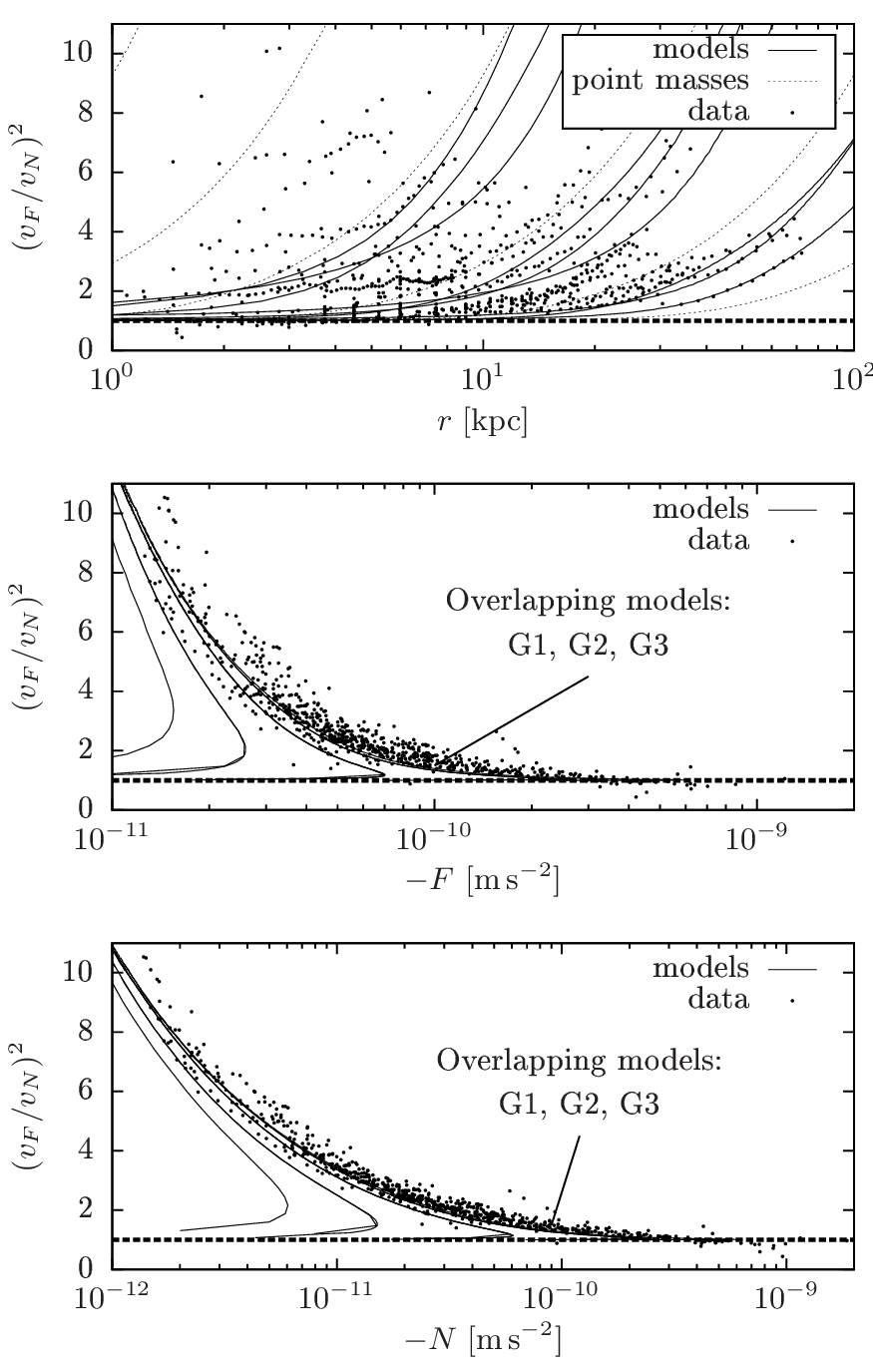}
  \caption{
    Discrepancies between Newtonian expectations and SRG as measured with the
    squared ratio of their orbital velocities, $v_N$ and $v_F$, respectively.
    The top panel displays the radial profile, whereas the middle and bottom
    panels the dependence of this ratio on the SRG and Newtonian forces.
    The solid lines represent models G0--G8.
    The curves of the low-surface-density galaxies appear on the leftmost region
    of the middle and bottom panels, whereas they move to the right as
    $\sigma_0$ increases.
    The dotted lines represent point masses with
    $m = 10^{7}M_\odot, 10^{8}M_\odot, ..., 10^{12}M_\odot$ (left to right in
    the top panel, they coincide in the middle and bottom ones).
    The larger the mass, the farther away the deviations become apparent.
    Dots are observational data from refs.~\cite{McGaugh04, FamaeyMcGaugh12},
    and the thick dashed line highlights $(v_F/v_N)^2 = 1$.}
  \label{fig:massdiscr}
\end{figure}
Figure~\ref{fig:massdiscr} shows the expected squared ratio between the SRG and
Newtonian velocities $(v_F/v_N)^2$ for the disk galaxy models G0--G8 (solid
lines).
Observational data for the squared ratio between the measured and Newtonian
(implied by the baryonic matter) velocities are overplotted for a sample of disk
galaxies (dots) \cite{McGaugh04, FamaeyMcGaugh12}.
We also display the curves for a selection of point masses with
$m = 10^{7}M_\odot, 10^{8}M_\odot, ..., 10^{12}M_\odot$ -- placed in between
parallel plane boundaries with $h = \sqrt{Gm/b}$ (dotted lines) -- in order to
capture the profiles of Eqs.~(\ref{eq:ratio-r}), (\ref{eq:ratio-f}), and
(\ref{eq:ratio-n}).
The middle and bottom panels show that there is no mass discrepancy when the SRG
and Newtonian accelerations are greater than
$b\sim 1.2 \cdot 10^{-10}\,\mathrm{m\,s}^{-2}$.
In contrast, the mass discrepancy increases towards the other limit, with the
dependencies in good agreement with Eqs.~(\ref{eq:ratio-r}), (\ref{eq:ratio-f}),
and (\ref{eq:ratio-n}).
A larger value of $m$ relocates the radial profiles to the right (top panel),
whereas the relation between mass discrepancy and force is independent of $m$,
and thus all the dotted lines coincide (middle and bottom panels).
The larger the surface density of a disk galaxy the closer it is to the
point-mass curves (middle and bottom panels), with the models G1--G3 being
indistinguishable from the underlying dotted lines.
These models fit adequately well the observed mass discrepancy data as a
function of either the Newtonian or the SRG force.
We note that for small $\sigma_0$ the value of $b$ is never reached by $F$, as
also seen in figure~\ref{fig:rfvlog}.

\subsection{Velocity dispersions in the Milky Way}

In SRG disk galaxies the vertical scale length $h$ is large enough to allow the
motion of stars perpendicularly to the galactic disk and to allow the presence
of non-flat galactic subsystems, like the stellar halo or the system of globular
clusters.
   
The steady-state ($\partial/\partial t = 0$) Jeans equation for the momentum is
\begin{equation}
  \nabla\cdot(n\mathbf w) = n\mathbf F\,,
  \label{eq:jeans}
\end{equation}
where $n$ is the stellar number density, $\mathbf F$ is the SRG acceleration,
and $w_{ij} = \left<v_iv_j\right>$ is a tensor with the brackets $\left<\right>$
denoting averages over velocity space.
In cylindrical coordinates, by assuming
$\left<v_r\right> = \left<v_z\right> = \left<v_\phi\right> = 0$ we obtain
$w_{rr} = \big<v_r^2\big> = \sigma_r^2$,
$w_{\phi\phi} = \big<v_\phi^2\big> = \sigma_\phi^2$,
$w_{zz} = \big<v_z^2\big> = \sigma_z^2$, and
$w_{rz} = \left<v_rv_z\right> = \sigma_{rz}^2$, where $\sigma$ here denotes the
velocity dispersion.
Under the assumption of axisymmetry, we can rewrite the radial component of the
momentum equation as:
\begin{equation}
  \frac{1}{r}\frac{\partial}{\partial r}\left(rn\sigma_r^2\right)
    + \frac{\partial}{\partial z}\left(n\sigma_{rz}^2\right)
    - \frac{n\sigma_\phi^2}{r} = nF\,.
  \label{eq:jeansrad}
\end{equation}
At large distances, $r \gg h$, and near $z = 0$, we may approximate the force as
$F = -GM/hr = -v_\mathrm{f}^2/r$.
Then, by multiplying Eq.~(\ref{eq:jeansrad}) by $r/n\sigma_r^2$, we obtain
\begin{equation}
  \frac{d\ln(n\sigma_r^2)}{d\ln r} = - \frac{v_\mathrm{f}^2}{\sigma_r^2}
    + \frac{\sigma_\phi^2}{\sigma_r^2} -\frac{3}{2} \pm\frac{1}{2}
    + \frac{\sigma_z^2}{2\sigma_r^2} \mp \frac{\sigma_z^2}{2\sigma_r^2}
    \equiv -p\,.
\label{eq:p}
\end{equation}
In the last expression we have taken the two extreme possibilities for
$d\sigma_{rz}^2/dz$, i.e. either equal to zero (upper signs) or equal to
$\left(\sigma_r^2 - \sigma_z^2\right)/r$ (lower signs), see
ref.~\cite{BinneyTremaine08}.
In Newtonian gravity Eq.~(\ref{eq:p}) has an identical form, with the Newtonian
velocity $v_N^2=-rN$ in the place of $v_\mathrm{f}^2$.

We can apply Eq.~(\ref{eq:p}) to observations of the velocity dispersion profile
of the Milky Way stellar halo.
A fit to these data suggests that $\sigma_r$ is approximately
constant at distances $20$--$80\,\mathrm{kpc}$ near the disk plane
\cite{Gnedin+10}.
In the notation of ref.~\cite{Gnedin+10}, $\sigma_r$ is the component of the
velocity dispersion along the radius in spherical coordinates, rather than in
the cylindrical coordinates we use here.
The two radial coordinates are clearly equal on the disk plane and thus we
consider $\sigma_r$ not to vary with radius.
Since the anisotropy parameter $\beta$ is assumed to be constant, we may also
take $\sigma_\phi$ and $\sigma_z$ as constants.
Then Eq.~(\ref{eq:p}) can be integrated to give
\begin{equation}
  n(r) = Ar^{-p}\,,
\end{equation}
where $p$ and $A$ are constants.
In order for SRG to be consistent with observations the predicted scaling of the
number density has to match the profile of the Milky Way.

Observational data indicate that, to zeroth order, $v_\mathrm{f}$ is roughly in
the range $1.7\sigma_r$ to $2\sigma_r$ \cite{Gnedin+10}.
Thus the first term of $p$, $v_\mathrm{f}^2/\sigma_r^2$, takes a value between
$3$ and $4$.
Moreover, the approximate value of the anisotropy parameter, $\beta \simeq 0.4$
\cite{Gnedin+10}, indicates that the tangential and radial velocity dispersion
components are roughly equal; in our case this means
$(\sigma_\phi^2 + \sigma_z^2)/\sigma_r^2 \simeq 1.2$.
If we further consider $\sigma_\phi \simeq \sigma_z$ \cite{BinneyTremaine08} we
have $\sigma_\phi^2/\sigma_r^2 \simeq \sigma_z^2/\sigma_r^2 \simeq 0.6$.
By combining the terms together we find the zeroth order approximation for $p$
in the two limits, which is in the range $3.4$ and $4.4$ for the upper signs,
and between $3.8$ and $4.8$ for the lower signs.
These agree with the value $\gamma_\mathrm{tr} \simeq 4$ suggested in
ref.~\cite{Gnedin+10}, which is a parameter equivalent to $p$.
Therefore, SRG appears to be consistent with the velocity dispersions observed
in the Milky Way stellar halo.

\section{Solving the RG field equation}
  \label{sec:rgpoisson}

\subsection{Application to a model disk galaxy}

\begin{figure}
  \centering
  \includegraphics[width=0.7\textwidth]{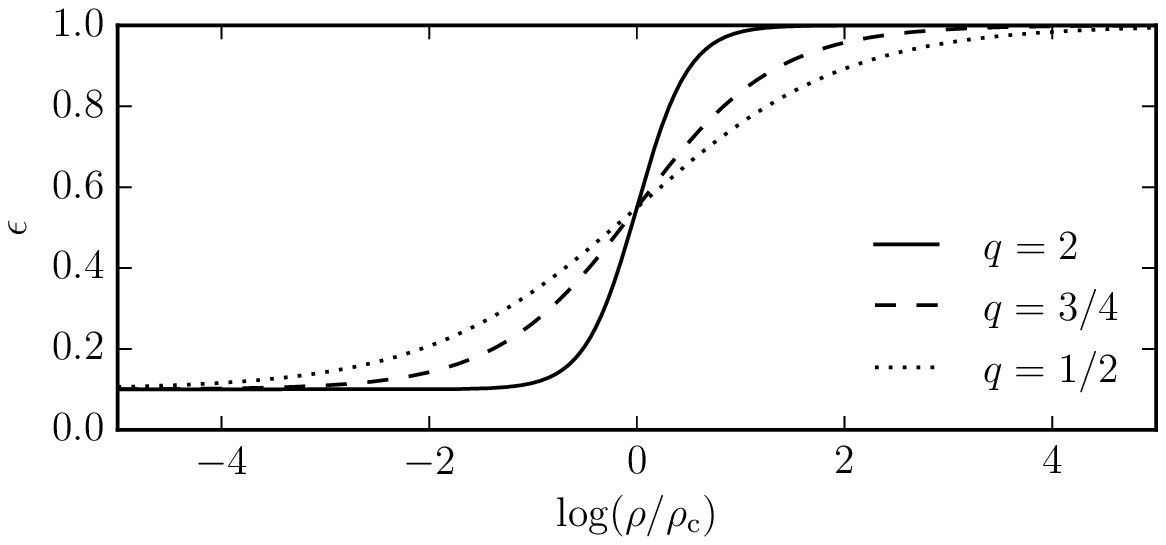}
  \caption{
    A simple assumption for the profile of the gravitational permittivity:
    Eq.~(\ref{eq:epsfunction}) is plotted as a function of
    $\rho/\rho_\mathrm{c}$ for $\epsilon_0 = 0.1$ and three different values of
    $q$.}
  \label{fig:epsilon}
\end{figure}
Having tested the analytical predictions of the SRG approximation, we proceed
here to solve the full RG Poisson equation for a disk galaxy model from
Table~\ref{tab:spirals}.
We consider a time-independent density distribution,
$\rho(r,z) = [\sigma_0/(2z_0)]\exp(-r/r_0)\exp(-z/z_0)$ (adopting the parameters
of model G0), and we compute $\Phi$ by integrating Eq.~(\ref{eq:poisson}) with a
relaxation numerical method.
For the permittivity we assume a smooth step function from $\epsilon_0$ to $1$,
\begin{equation}
  \epsilon(\rho) = \epsilon_0 + (1 - \epsilon_0)
    \frac{1}{2}\left\{\tanh\left[\log\left(
    \frac{\rho}{\rho_\mathrm{c}}\right)^q\right] + 1\right\}\,,
  \label{eq:epsfunction}
\end{equation}
where $q$ is a parameter that controls the steepness of the transition, see
figure~\ref{fig:epsilon}.

\begin{figure}
  \centering
  \includegraphics[width=0.8\textwidth]{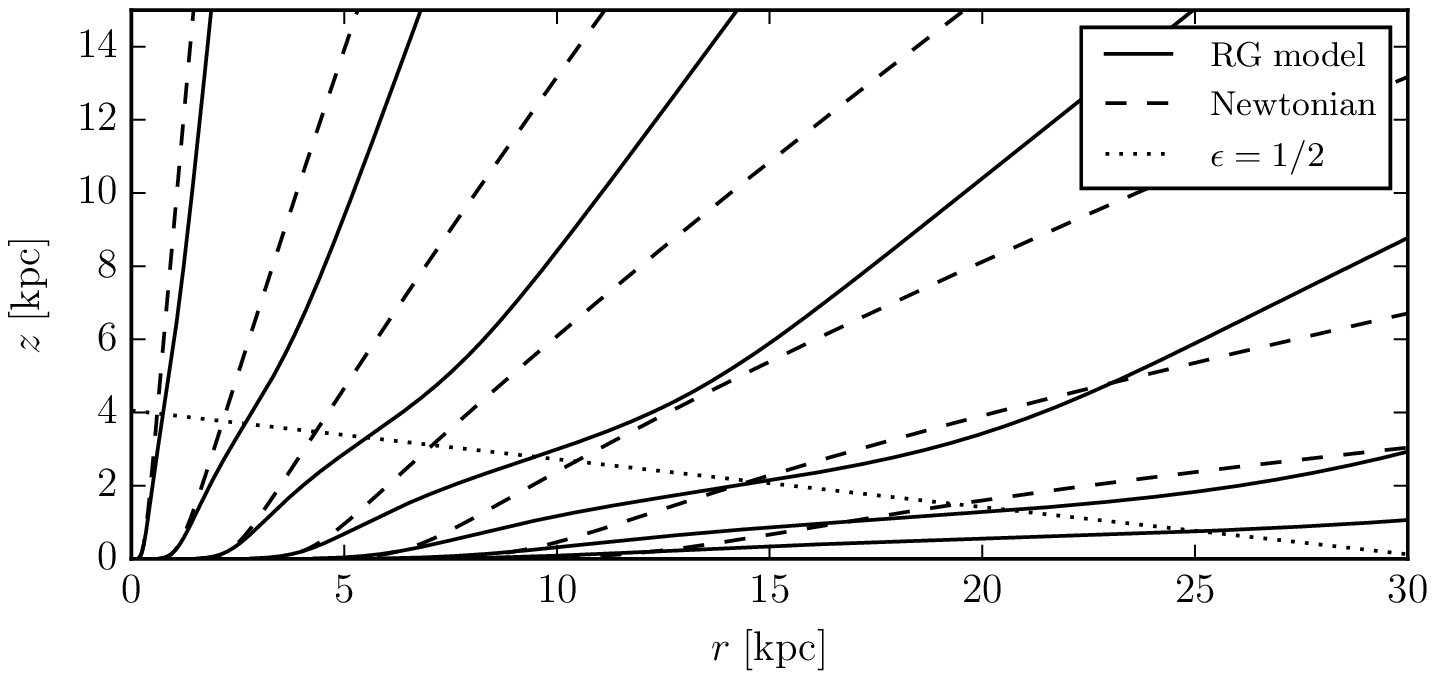}
  \includegraphics[width=0.8\textwidth]{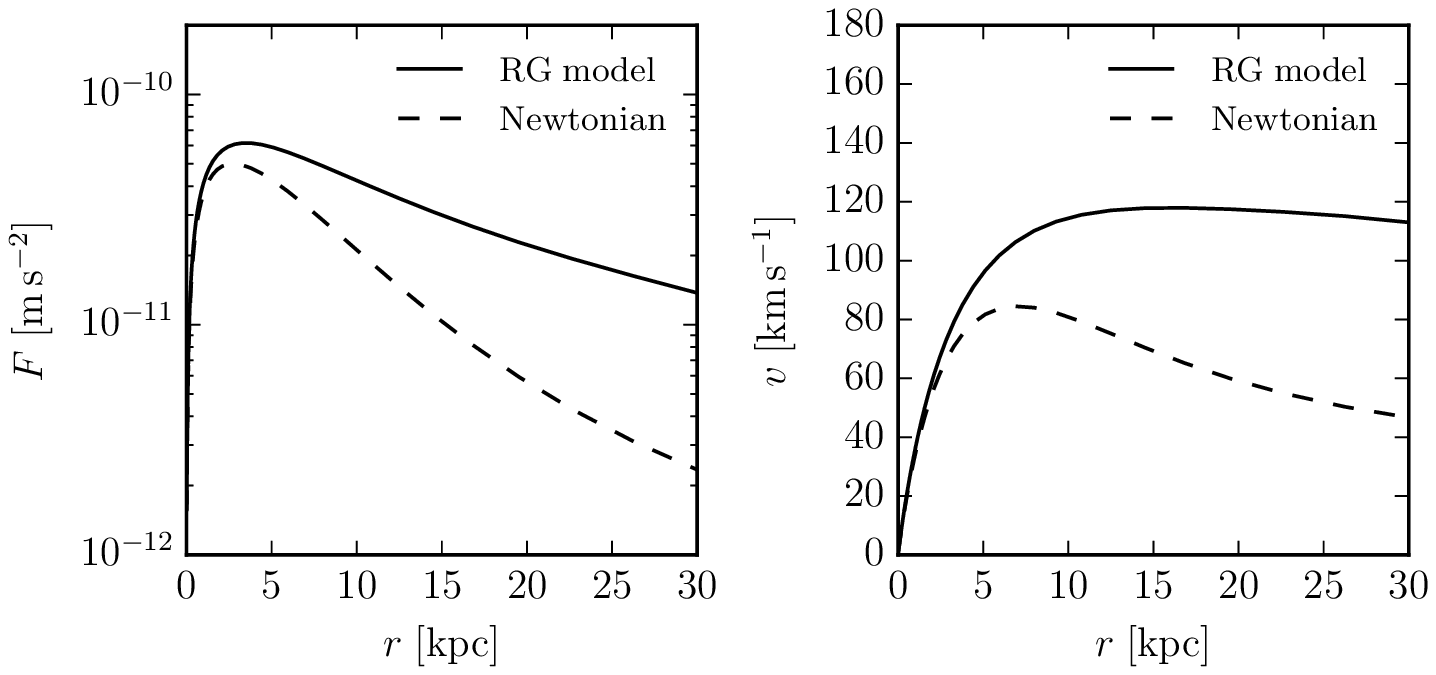}
  \caption{
    Top panel: Gravitational field lines (solid: RG, dashed: Newtonian gravity)
    for the disk galaxy model G0, as computed by Eqs.~(\ref{eq:poisson}) and
    (\ref{eq:epsfunction}) for the parameters $q = 1/2$,
    $\epsilon_0 = 1/6$, and $\rho_\mathrm{c} = 10^{-4}\sigma_0/2z_0$.
    The dotted line shows the contour $\epsilon = 1/2$.
    Bottom panels: Acceleration (left) and orbital velocity (right) on the
    midplane (compare with SRG in figure~\ref{fig:galaxy}).}
  \label{fig:fieldlines}
\end{figure}
The top panel of figure~\ref{fig:fieldlines} illustrates the gravitational field
lines of model G0 in the RG (solid) and Newtonian (dashed) regimes.
The refraction of the RG field is evident throughout the domain, with the effect
being stronger at the outer radii.
The two breaks seen along each field line, above and below the contour of
$\epsilon$ (dotted line), mark the transition from $\epsilon = 1$ to
$\epsilon_0$.
For the rotation curve, the most relevant feature is the redirection of the
gravitational flux that occurs at lower heights, $z \lesssim 2$\,kpc, and tends
to align the field lines parallely to the disk plane.

In the bottom panels of figure~\ref{fig:fieldlines} we plot the radial profiles
of the acceleration (left) and the rotation velocity (right) along the midplane.
A comparison with figure~\ref{fig:galaxy} confirms the validity of SRG as a
powerful tool to approximate the phenomenology of disk galaxies.
The slightly smaller $v_\mathrm{f}$ seen in figure~\ref{fig:fieldlines} is due
to the non-zero value of $\epsilon_0$ (as compared to $\epsilon_0 = 0$ for SRG,
figure~\ref{fig:galaxy}), which results in a less efficient reorientation of the
field lines along the midplane.
Notably, even though the dynamical behavior in the equatorial region
$r \lesssim 10\,\mathrm{kpc}$ strongly deviates from the Newtonian expectation,
the permittivity there is $\epsilon = 1$.
This result suggests that local experiments cannot detect any discrepancies from
standard gravity despite the presence of a strong background field of
non-Newtonian origin.

\subsection{Rotation curves of NGC1560 and NGC6946}
  \label{sec:NGCrotcurves}

\begin{figure}
  \centering
  \includegraphics[width=0.85\textwidth]{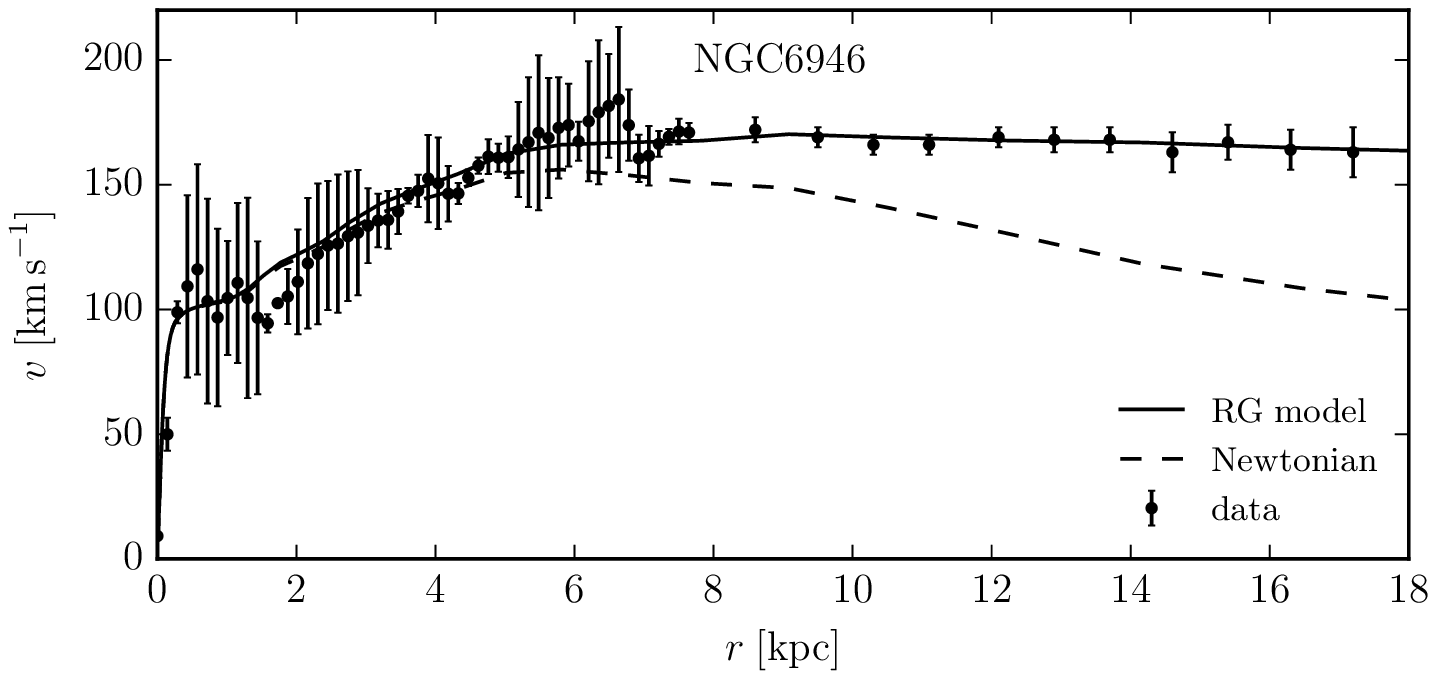}
  \includegraphics[width=0.85\textwidth]{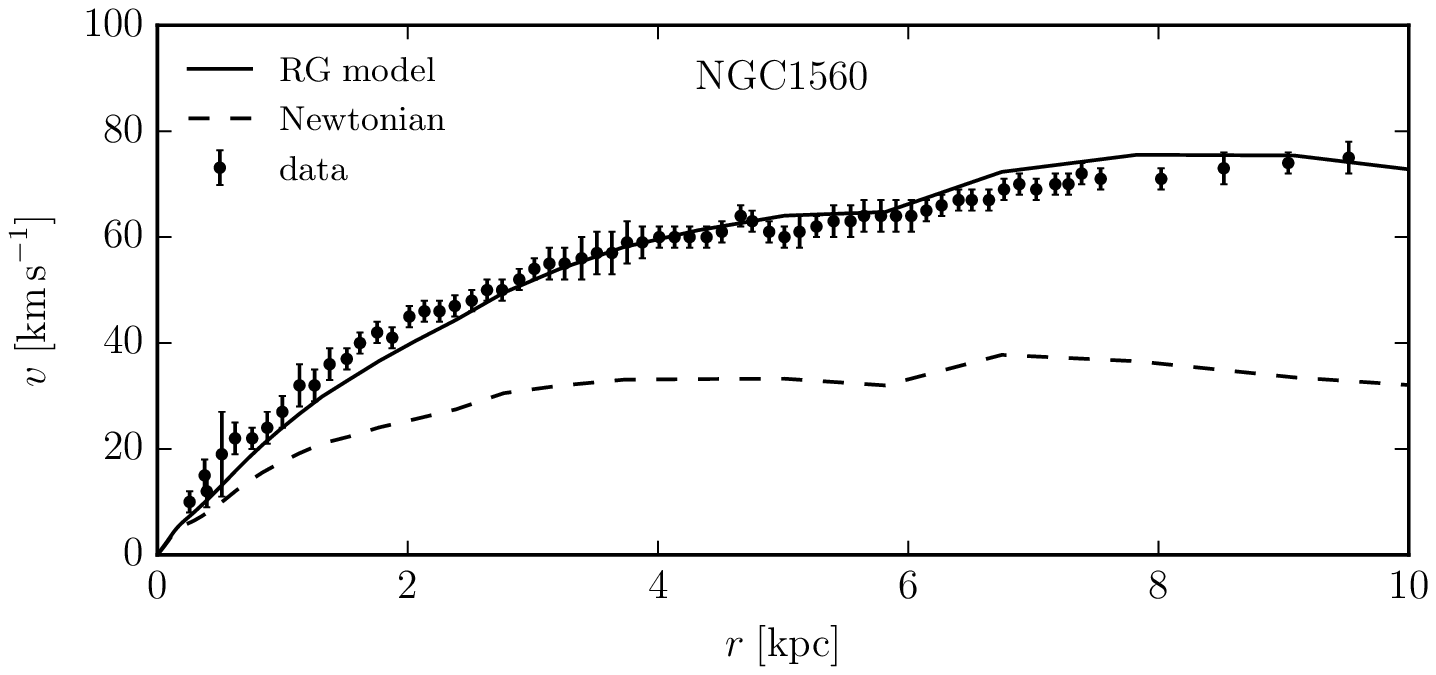}
  \caption{
    Rotation curves for the high-surface-brightness disk galaxy
    NGC6946 (top panel) and the low-surface-brightness disk galaxy NGC1560
    (bottom panel).
    The solid and dashed lines are derived by solving the RG and Newtonian
    Poisson equations, respectively.
    The NGC6946 data points are H$\alpha$ and HI observations from
    refs.~\cite{Daigle+06, Boomsma+08} and the NGC1560 data are HI observations
    from ref.~\cite{Gentile+10}.
    A Hubble constant $H_0 = 70$\,km\,s$^{-1}$\,Mpc$^{-1}$ is assumed.}
  \label{fig:obsgalaxies}
\end{figure}
As specific examples we model the observed rotation curves of two real disk
galaxies, the high-surface-brightness NGC6946 and the low-surface-brightness
NGC1560, using the surface density distributions given in
ref.~\cite{FamaeyMcGaugh12}.
We complement the radial dependency with an exponential vertical profile,
$\exp(-z/z_0)$, assuming $z_0 = 0.6\,\mathrm{kpc}$ (NGC6946) and
$z_0 = 0.2\,\mathrm{kpc}$ (NGC1550) \cite{Kregel+02}, and we rescale the overall
$\rho$ such that the Newtonian profiles match those shown in
ref.~\cite{FamaeyMcGaugh12} (this is equivalent to readjusting the mass-to-light
ratio).
For the permittivity we consider Eq.~(\ref{eq:epsfunction}) with $q = 3/4$,
$\epsilon_0 = 0.25$ and $0.20$, and critical densities
$\rho_\mathrm{c} = 10^{-24}\,\mathrm{g}\,\mathrm{cm}^{-3}$ and
$10^{-27}\,\mathrm{g}\,\mathrm{cm}^{-3}$, respectively.
The panels of figure~\ref{fig:obsgalaxies} show the RG predictions for each
galaxy (solid lines), which are found in agreement with the observed rotation
curves (points) \cite{Daigle+06, Boomsma+08, Gentile+10}.
We note that the smoothness of the curve and the intensity of its features may
be better captured by assuming a radially-dependent scale height.
However, we refrain from improving or quantifying the fit because the assumed
expression for $\epsilon$ consists of a simplistic hypothesis, only appropriate
for demonstrating the proof of concept.
Further development of RG is needed for a more detailed analysis.

\subsection{Intracluster gas temperature of A1991 and A1795}
  \label{sec:clusters}

\begin{figure}
  \centering
  \includegraphics[width=0.85\textwidth]{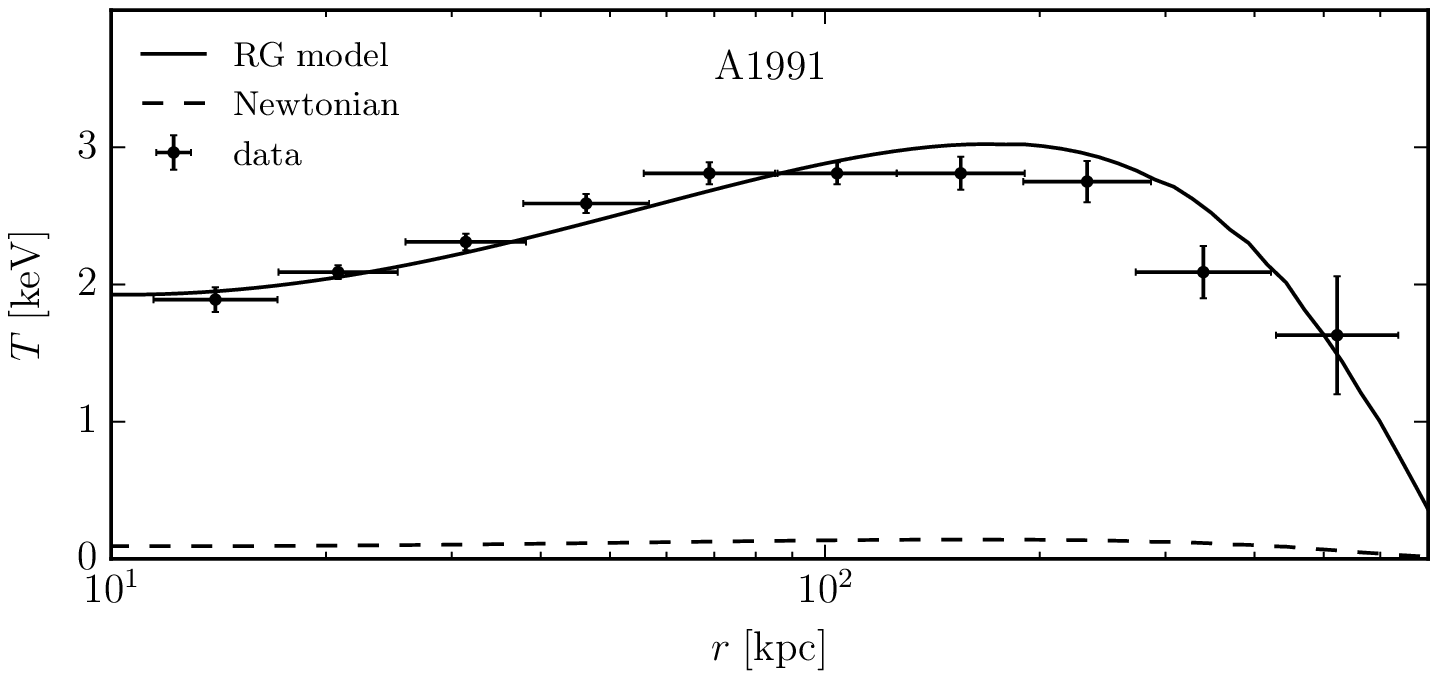}
  \includegraphics[width=0.85\textwidth]{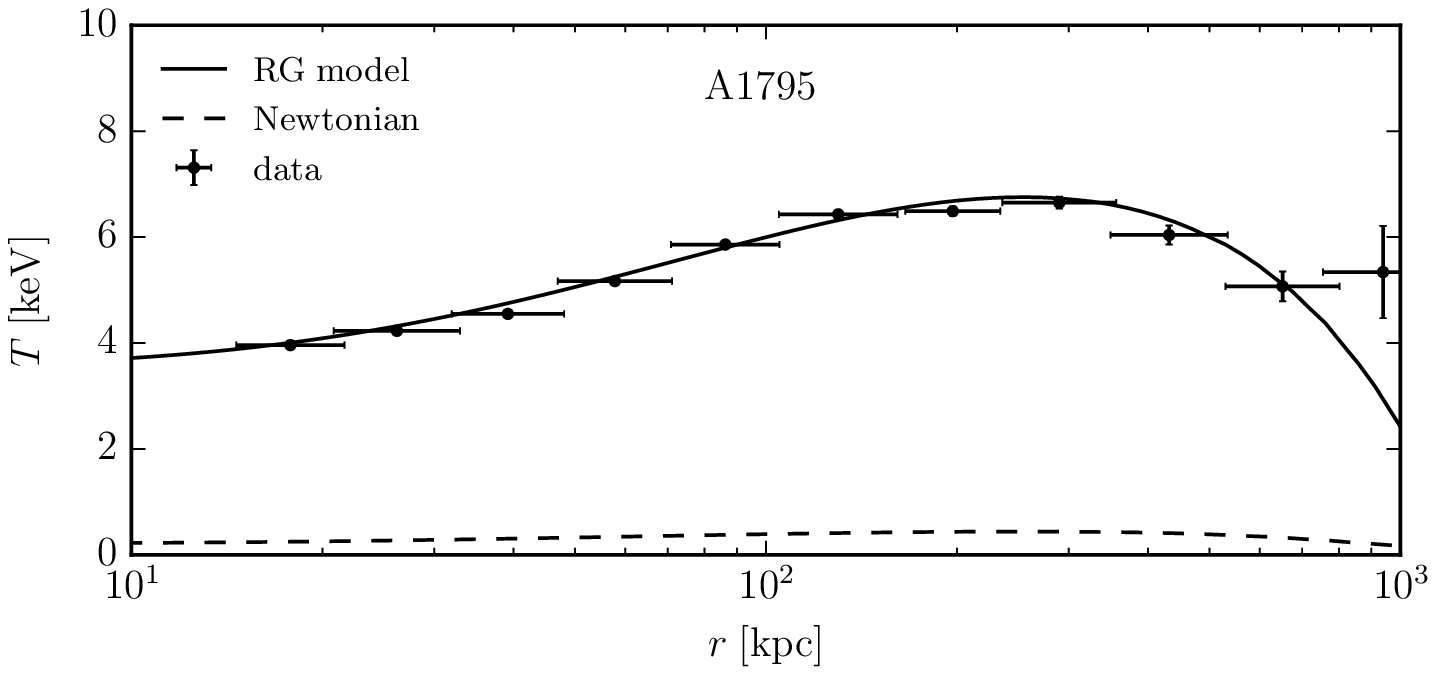}
  \caption{
    Emission-weighted projected temperature profiles of the intracluster gas in
    two galaxy clusters, A1991 (top panel) and A1795 (bottom panel).
    The solid lines correspond to the RG models, the dashed lines to the
    Newtonian expectation, and the data points are \textit{Chandra} observations
    from ref.~\cite{Vikhlinin+06}.
    A Hubble constant $H_0 = 71$\,km\,s$^{-1}$\,Mpc$^{-1}$ is assumed.}
  \label{fig:obsclusters}
\end{figure}
Finally, we apply the RG field equation to two galaxy clusters of different gas
temperatures, A1991 and A1795 \cite{Vikhlinin+06}.
We set up a spherically-symmetric density distribution according to the observed
gas profiles (truncated at $r_{500}$ \cite{Vikhlinin+06}, a distance larger than
the plotted range), and we add a stellar component of mass
$5\cdot10^{10}M_\odot$ to the inner radii, $R < 10\,$kpc, of A1991.
The permittivity is assumed to follow Eq.~(\ref{eq:epsfunction}), with the
values here being $q = 2$, $\epsilon_0 = 0.045$ and $0.065$, respectively, and
$\rho_\mathrm{c} = 10^{-24}\,\mathrm{g}\,\mathrm{cm}^{-3}$.
The RG field is computed first, and then the gas temperature is calculated based
on the hydrostatic equilibrium and the ideal-gas equation of state.
Figure~\ref{fig:obsclusters} shows that the emission-weighted projected
temperatures that are predicted by RG (solid lines) are in agreement with
\textit{Chandra} measurements (points) \cite{Vikhlinin+06}.
The fit can be further improved by including the mass profiles of the galaxies
and also by assuming more sophisticated permittivity functions beyond the
speculative expression of Eq.~(\ref{eq:epsfunction}).

\section{Discussion}
  \label{sec:discussion}

\subsection{Disk scale heights $z_0$ and $h$}
  \label{sec:assumptions}

\begin{figure}
  \centering
  \includegraphics[width=0.65\textwidth]{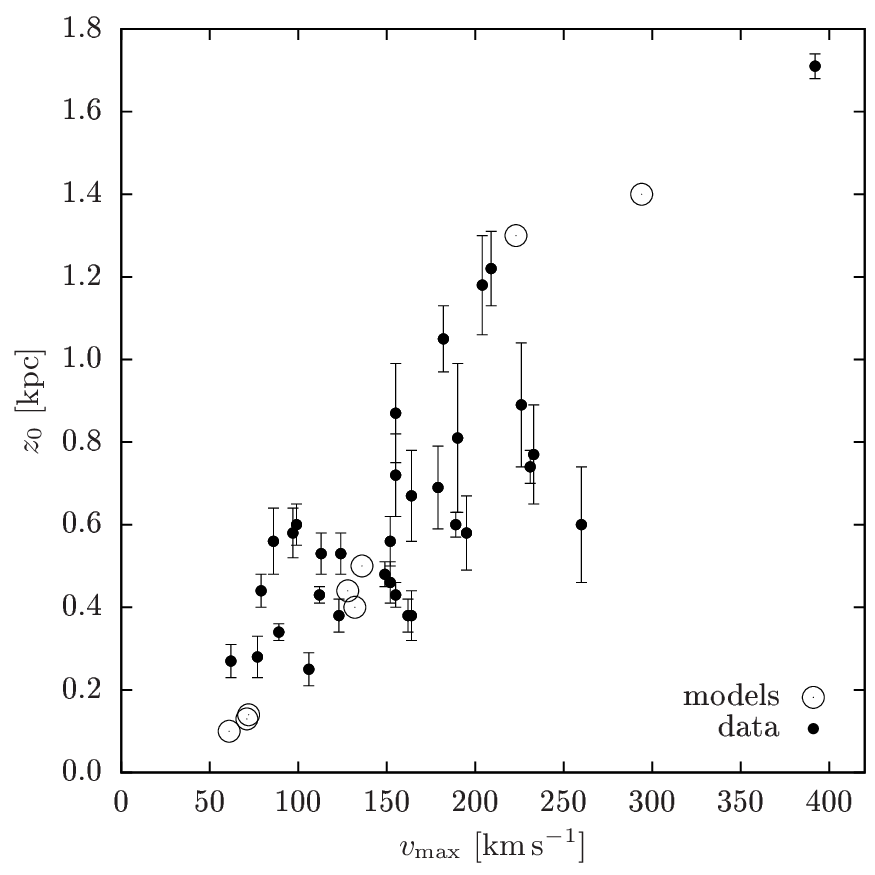}
  \caption{
    The scale height as a function of the maximum orbital velocity.
    The dots are galaxies from ref.~\cite{Kregel+02} and the open circles
    represent the models G0--G8 assuming $z_0 = 0.1h$.}
  \label{fig:z0v}
\end{figure}
Theoretical arguments as well as observational evidence suggest that the disk
scale height $z_0$ is independent of radius \cite{vanderKruitSearle81}.
A potential correlation of $h$ with $z_0$, i.e. $h \propto z_0$, would connect
$h$ with the contours of density, thus supporting the RG assumptions and the
simplified equivalent framework of SRG.
To investigate the hypothesis $h \propto z_0$, we show in figure~\ref{fig:z0v}
observational data (dots) for the scale height $z_0$ as a function of the
maximum orbital velocity \cite{Kregel+02}.
We arbitrarily choose $z_0 = 0.1h$ and overplot the corresponding points (open
circles) of the models listed in Table~\ref{tab:spirals}.
A visual inspection of this plot suggests that the assumption $h \propto z_0$ is
a reasonable first approximation.

\subsection{Relation to dark matter}
  \label{sec:dm}

In the case of disk galaxies the radial dependence of the Newtonian and RG
gravitational fields at radii $r \gg h$ agree with each other if, in Newtonian
gravity, we assume the presence of dark matter with mass $M_\mathrm{DM}$:
\begin{equation}
  \frac{G(M_\mathrm{DM} + m)}{r^2} = \frac{Gm}{hr}\,.
\end{equation}
The required amount of dark matter is thus $M_\mathrm{DM} = m(r/h-1)$ and its
density profile is $\rho_\mathrm{DM} =
\mathrm{d}M_\mathrm{DM}/(4\pi r^2\mathrm{d}r) \simeq m/4\pi hr^2$.
The power law behavior $\rho_\mathrm{DM} \propto r^{-2}$ agrees with the NFW
density profile of DM halos \cite{Navarro+97} in the region where
$r \sim r_\mathrm{s}$ ($r_\mathrm{s}$ is the scale radius in the notation of
ref.~\cite{Navarro+97}) that is inferred from the dynamics of real galaxies when
interpreted with Newtonian gravity.

A similar argument can be applied to gravitational lensing.
In General Relativity the deflection angle of a light ray passing at distance
$r$ from a point mass $m$ is $\theta = 4Gm/rc^2$.
To zeroth order, the modification within the RG framework is
\begin{equation}
  \theta = \frac{4Gm}{rc^2}\frac{1}{\epsilon_0}\,,
  \label{eq:darkmatter}
\end{equation}
where $m$ is ordinary matter that, because of the permittivity of the medium,
originates a gravitational field a factor $1/\epsilon_0$ stronger.
We thus expect that the gas density profile in the outskirts of galaxy clusters 
is similar to the DM density profile inferred in Newtonian gravity, even if its
density is approximately an order of magnitude smaller: this feature is indeed
observed in X-ray clusters \cite{Vikhlinin+06}.

\begin{figure}
  \centering
  \includegraphics[width=\textwidth]{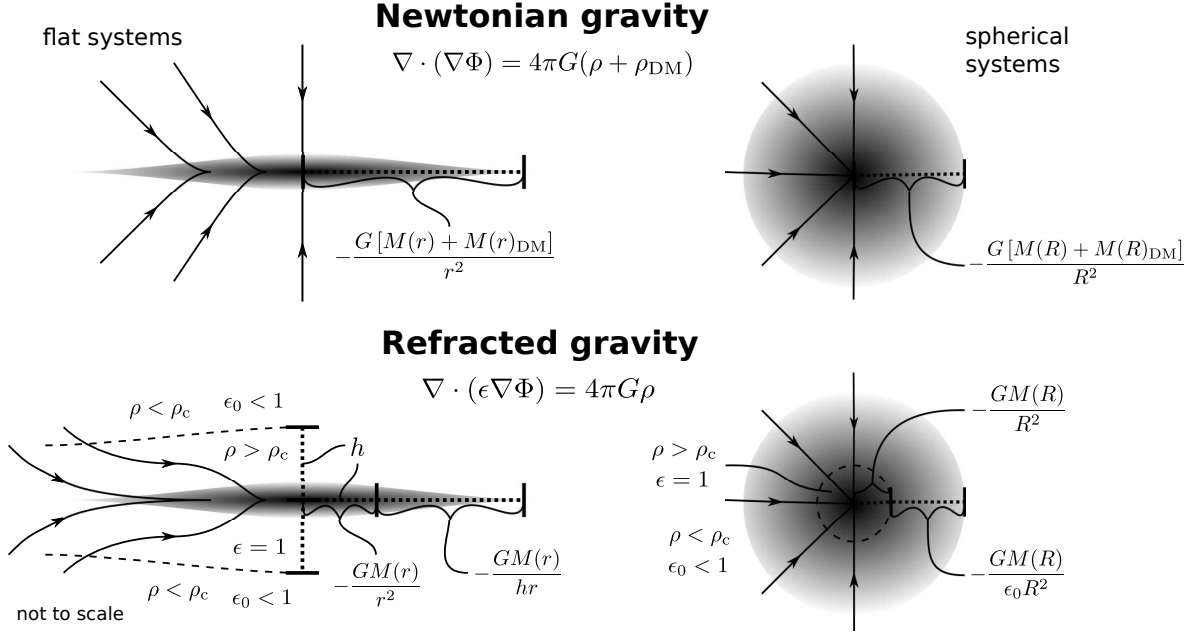}
  \caption{
    Schematic representation of the gravitational field lines and the
    approximate radial dependence of the acceleration in flat and spherical
    systems (top: Newtonian gravity with DM; bottom: RG).
    The functions $M(r)$ and $M(R)$ refer to the total enclosed mass within the
    cylindrical and spherical radius, respectively.}
  \label{fig:sketch}
\end{figure}
Figure~\ref{fig:sketch} shows a schematic illustration of the radial dependence
of the gravitational force in flat (left) and spherical systems (right), as
described by Newtonian gravity in the presence of DM (top) and by the RG
framework (bottom).

\subsection{Relation to MOND}
  \label{sec:mond}

MOND states that below a critical acceleration (denoted with $a_0$) the force
becomes
\begin{equation}
  F_\mathrm{MOND} = \sqrt{\frac{Gm}{r^2}a_0}\,.
\end{equation}
By substituting $a_0$ with $b$ from Eq.~(\ref{eq:b}) we obtain:
\begin{equation}
  F_\mathrm{MOND} = \sqrt{Nb} = \sqrt{\frac{Gm}{r^2}b} = \frac{Gm}{rh}\,;
\end{equation}
namely, it has the same profile as RG on the midplane of a flat object at
distances $r \gg h$.
Therefore, we expect that a significant portion of the observational successes
of MOND concerning disk galaxies also belongs to RG.
However, the differences between RG and MOND are fundamental.

According to the formulation of ref.~\cite{BekensteinMilgrom84}, the Poisson
equation in MOND reads
\begin{equation}
  \nabla\cdot(\mu\nabla\Phi) = 4\pi G\rho\,,
\end{equation}
where $\mu = \mu(|\nabla\Phi|/a_0)$.
This formula is only formally similar to Eq.~(\ref{eq:poisson}) of RG, because
in MOND the factor $\mu$ depends on $\nabla\Phi$ itself (i.e. the field), rather
than on the mass density (i.e. the source, as in the case of $\epsilon$).
Moreover, MOND exhibits non-Newtonian behavior only in regions where the local
acceleration drops below the critical value $a < a_0$, but this is not the case
for RG and its corresponding critical parameter $\rho_\mathrm{c}$.
As an example consider the midplane of disk galaxies, where even though the
density is above the critical value ($\rho > \rho_\mathrm{c}$ and thus
$\epsilon = 1$) RG predicts strong deviations from Newtonian gravity (see
figure~\ref{fig:fieldlines}).
This is due to the redirection of the gravitational flux that occurs non-locally
and provides a stronger background field.
Nonetheless, since $\epsilon = 1$ there, local experiments would be in full
agreement with standard gravity.

We further highlight the fundamental differences by comparing the predictions of
RG and MOND for spherical and flat sources.
First, consider a point mass $m$ within a background density
$\rho > \rho_\mathrm{c}$.
RG gives the exact Newtonian behavior for any radius $R$, i.e.
$F = N \propto R^{-2}$.
Even though $F_\mathrm{MOND}$ follows this profile close to the source, the
radial dependence becomes $F_\mathrm{MOND}\propto R^{-1}$ at large distances
where the acceleration falls below $a_0$.
Consider now the case where the point source is in empty space, for which MOND
exhibits the same phenomenology as the example above.
In RG the force now is $F = N/\epsilon_0 \propto R^{-2}$ (for any radius), which
implies that a gravitational source surrounded by empty space appears to be more
massive due to the permittivity of vacuum $\epsilon_0 < 1$, but the power law
remains Newtonian.
In principle, RG and MOND can be distinguished observationally by measuring the
field in spherical systems, such as globular or galaxy clusters, at large
distances where the background density and/or the acceleration drop below the
critical value.

On the other hand, both MOND and RG predict almost the same radial dependence on
the midplane of flat objects: the gravitational field is proportional to
$r^{-2}$ at inner radii and to $r^{-1}$ in the outskirts.
However, above and below the disk the predictions can be different.
Depending on the height and how the field lines are refracted (location and
shape of $\rho_\mathrm{c}$), RG is expected to give a Newtonian vertical field
within $z = \pm h$.
In contrast MOND suggests a strong acceleration, which is always in the MONDian
regime far away from the source.
A detailed analysis of the rotation curves and the stellar velocity dispersion
profiles of disk galaxies from the DiskMass Survey \cite{Angus+15} has recently
shown that MOND requires a relation between scale lengths and scale heights
substantially flatter than observed: in MOND, once the mass-to-light ratio is
set by the rotation curve, the vertical component of the gravitational force is
larger than in Newtonian gravity.
This effect propagates into the thickness of a disk galaxy in MOND: the
thickness should be a factor of 2 smaller than the thickness estimated from the
observation of edge-on galaxies \cite{Angus+15}.
We expect that RG might not suffer from this problem.

Similarly, in some dwarf spheroidal galaxies MOND requires radial velocity
dispersion profiles larger than observed \cite{Angus08, Angus+14}, a fact that
might again be naturally solved by RG.
A detailed analysis of all these topics within the RG framework is clearly
needed.

\section{Other applications}
  \label{sec:applications}

The following subsections discuss the relevance of the RG framework to other
systems that exhibit mass discrepancies.
We plan to investigate quantitatively the predictions of RG for such systems in
future work.

\subsection{Globular clusters}
  \label{sec:globular}

Since globular clusters are spherically symmetric systems, RG does not predict
any redirection of the gravitational field lines.
In addition, as long as the background density is larger than $\rho_\mathrm{c}$,
no substantial deviations from Newtonian dynamics are expected.
This behavior is compatible with the numerous observations that confirm the
validity of Newtonian gravity, even though the acceleration over a part of the
globular cluster volumes is below $a_0$ \cite[e.g.,][]{Baumgardt+05,
Baumgardt+09, Jordi+09, Ibata+11a, Ibata+11b, Frank+12}.
However, there are systems that suggest a non-Newtonian behavior at large
distances \cite[e.g.,][]{Scarpa+07, Scarpa+10, Scarpa+11}, and these could
possibly be addressed with a change in $\epsilon$.
By combining measurements of the background density profile in Newtonian and
non-Newtonian clusters we can in principle constrain the critical values of the
density $\rho_\mathrm{c}$ and the permittivity $\epsilon$.

\subsection{Dwarf spheroidal galaxies}
  \label{sec:dwarf}

Dwarf spheroidal galaxies, whose baryonic masses are comparable to the mass of
globular clusters but have a much lower surface brightness, show evidence of
large mass discrepancies \cite[e.g.,][]{Walker+07, Strigari+08a, Strigari+08b}.
In RG this mass discrepancy is a consequence of the dwarf galaxy shapes, which
is significantly flatter than in globular clusters \cite[e.g.,][]
{vandenBergh08a, vandenBergh08b}.
In particular, we expect a departure from Newtonian gravity due to the
redirection of the gravitational flux along the dwarf midplane.
Because of the low surface brightness of dwarfs, RG suggests that the
non-Newtonian features are even stronger than in disk galaxies, as it can be
seen from figure~\ref{fig:rfvlog} and the corresponding discussion: the value of
$h$ is relatively smaller, and thus a non-Newtonian behavior is expected almost
within the entire volume of the dwarf ($r > h$).

\subsection{Elliptical galaxies}
  \label{sec:elliptical}

It has recently been suggested that the amount of extra mass required to explain
the dynamics of elliptical galaxies depends on the ellipticity of their shape
\cite{Deur13}.
Specifically, flatter elliptical galaxies seem to show larger mass discrepancies
than more roundish ones.
In the same context as globular clusters and dwarf spheroidal galaxies, this
feature arises naturally within the framework of RG: the flatter the shape of a
galaxy is, the stronger the field-redirection effect becomes.
Moreover, in RG the Tully-Fisher relation (Eq.~\ref{eq:tf}) can be derived only
for the case of disk galaxies, but it is not expected to hold for galaxies
without a flat morphology.
This conclusion seems to be supported by a recent study \cite{Simons+15} which
suggests that disk-like galaxies tightly follow the Tully-Fisher relation,
whereas those that scatter off are morphologically different.

\subsection{Disk galaxies}
  \label{sec:morespirals}

Since ordinary matter is the only source of gravity within the framework of RG,
the correlation between the features of the rotation curves and the features of
the surface brightness distribution, as observed in several disk galaxies (the
so-called ``Renzo's rule'') \cite{Sancisi04, Swaters+12}, is clearly expected.
Furthermore, the observed scaling relation $dv/dr \propto \sqrt{\rho_0}$ 
(where $\rho_0$ is the central density), holding in the inner regions of disk
galaxies \cite{Lelli+13}, also naturally arises in RG because its gravity
intensity reduces to the Newtonian intensity close to the galactic center.

\subsection{Galaxy groups and clusters}

In RG the evaluation of the gravitational interactions between galaxies, that
occur in groups and clusters, is complicated by the redirection of the lines of
the gravitational field originated by each galaxy faraway from the galaxy main
body, namely beyond the surface of constant critical density $\rho_c$.
As shown in the example of a disk galaxy in figure~\ref{fig:fieldlines}, the
shape of the iso-contour of $\rho_c$ determines the refraction angle and thus
the dependence of the gravitational force on the radial distance $r$ from the
galaxy center. 

If we overlook this complication and simply assume that the galactic
gravitational field of each galaxy decreases with $r^{-1}$ at large radii, the
net modification of RG, compared to standard gravity, is to increase the field
intensity by a factor $1/\epsilon_0$ due the gravitational permittivity of
low-density regions (see section~\ref{sec:spherical}).
Therefore, the extension of the baryonic Tully-Fisher relation (Eq.~\ref{eq:tf})
to galaxy groups and clusters becomes
\begin{equation}
  v_\mathrm{f}^4 = Gbf\frac{km}{\epsilon_0}\,,
  \label{eq:tfcluster}
\end{equation}
where $k$ is the number of galaxies (each with mass $m$) in the system and
$f < 1$ is a geometric factor that accounts for the fact that, due to the
anisotropic RG field geometry of disk galaxies, only a fraction of the $k$
members will cross the plane of a given disk galaxy where the force is
$\propto r^{-1}$.

\begin{figure}
  \centering
  \includegraphics[width=0.65\textwidth]{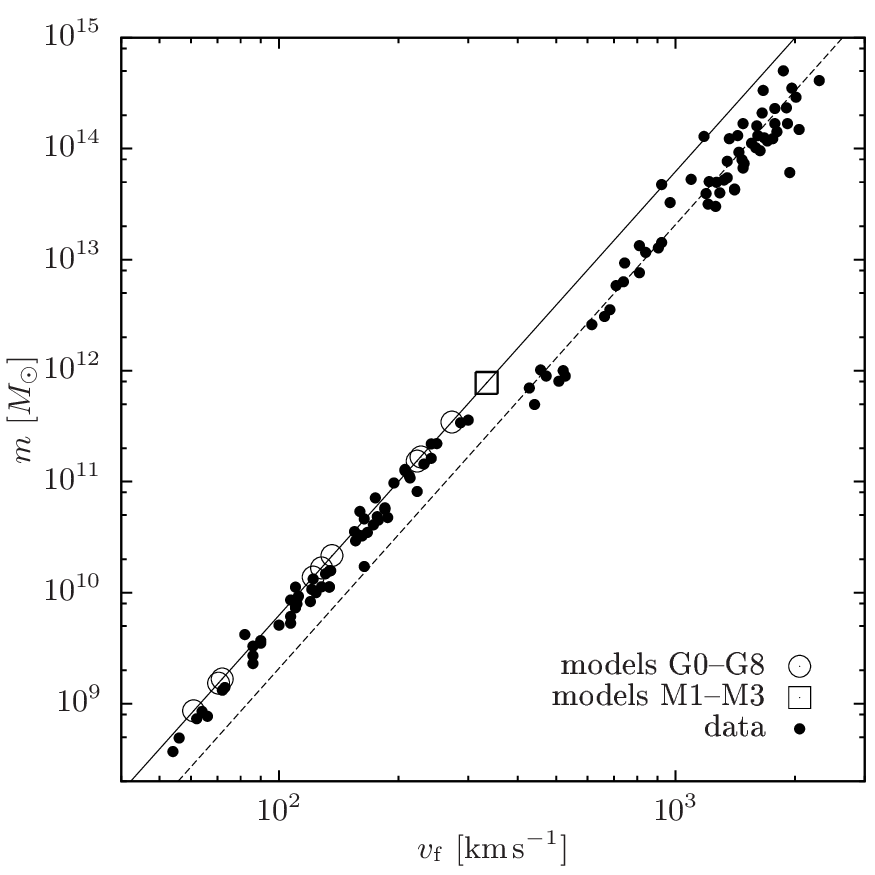}
  \caption{
    The baryonic Tully-Fisher relation -- same as the top panel of
    figure~\ref{fig:tf}, but at larger mass and velocity scales in order to
    include galaxy groups and galaxy clusters.
    Equation~(\ref{eq:tf}) is plotted with a solid line and
    Eq.~(\ref{eq:tfcluster}) is plotted with a dashed line for
    $f/\epsilon_0 = 1.3$.
    The additional observational data for groups and clusters that appear in the
    upper right part of the plot (dots) are taken from refs.~\cite{Sanders03,
    McGaugh15}.}
  \label{fig:tf2}
\end{figure}
Figure~\ref{fig:tf2} shows Eq.~(\ref{eq:tfcluster}) for
$f/\epsilon_0 = 1.3$ (dashed line), along with data points for galaxy groups and
clusters \cite{Sanders03, McGaugh15}.
The observed baryonic Tully-Fisher relation for such systems shows a smaller
normalization than the relation for galaxies, but its slope remains unaltered in
agreement with the approximate RG expectation.
In fact, this qualitative analysis suggests a procedure to estimate the critical
value of the density and the permittivity $\epsilon_0$.
In addition, Eq.~(\ref{eq:tfcluster}) implicitly assumes that the intracluster
medium (ICM) has a density below the critical value, as it was also indicated by
the fits in section~\ref{sec:clusters}.
This property could be used to set a lower limit to the value of
$\rho_\mathrm{c}$.

\section{Conclusion}
  \label{sec:conclusions}

In this paper we investigate an alternative approach to describe the dynamics of
galaxies and larger cosmic structures without requiring the existence of dark
matter.
Motivated by the Poisson equation of electrodynamics in matter, we propose that
the mass discrepancies originate from a gravitational permittivity that depends
on the local mass density and effectively refracts the field lines.
This effect can make the gravitational field significantly stronger than the
Newtonian one, with the expected deviations depending on both the total mass and
its distribution.

We show that this framework can successfully describe several observational
features of disk galaxies, such as the rotation curves and the baryonic
Tully-Fisher relation, as well as the basic phenomenology of galaxy groups and
clusters.
We also discuss the specific features of refracted gravity that might explain
why systems of similar luminous mass have significantly different dynamical
properties, for example globular clusters and dwarf galaxies, or disk and
elliptical galaxies.

It thus appears that refracted gravity might provide a promising approach to
describe the phenomenology of cosmic structures.
We conclude that it is worth developing the theoretical foundation of this model
in detail and investigating its observational expectations more rigorously.
This future work will assess whether refracted gravity can ultimately be
considered a valid theory or suffers from unsolvable failures.

\acknowledgments
We are grateful to Stacy McGaugh for providing us with the data points that
appear in several of our figures, as well as for his useful comments.
We would also like to thank Garry Angus and Justin Khoury for fruitful
discussions on several aspects of this work.
AD acknowledges partial support from the INFN grant InDark, the grant Progetti
di Ateneo TO Call 2012 0011 `Marco Polo' of the University of Torino and the
grant PRIN 2012 ``Fisica Astroparticellare Teorica'' of the Italian Ministry of
University and Research.

\bibliographystyle{JHEP}
\bibliography{paper}

\end{document}